\documentclass[conference,compsoc]{IEEEtran}

\usepackage{CJKutf8}
\pdfmapfile{+gbsnu.map}
\pdfmapfile{+gkaiu.map}
\usepackage{xcolor}
\usepackage{url}

\ifCLASSOPTIONcompsoc
  \usepackage[nocompress]{cite}
\else
  \usepackage{cite}
\fi
\newif\ifscaffold
\scaffoldtrue

\ifscaffold

\newcommand{\struct}[1]{{\color{blue!90!black}$\langle$\ttfamily\CJKfamily{gkai}#1$\rangle$}}
\newcommand{\ph}[1]{\struct{#1}} 

\else

\newcommand{\struct}[1]{}
\newcommand{\ph}[1]{}

\fi

\usepackage{amsmath}
\usepackage{amssymb}
\usepackage{amsthm}
\usepackage{bm}
\usepackage{xcolor}
\usepackage{tikz}
\usetikzlibrary{arrows.meta,fit,calc}
\usepackage{graphicx}
\usepackage{booktabs} 
\usepackage{yquant} 
\usepackage{hyperref}

\hypersetup{
  colorlinks=true,
  linkcolor=black!50!blue,
  citecolor=cyan!60!black,
  urlcolor=blue!70!black
}

\allowdisplaybreaks

\renewcommand\a{\alpha}
\renewcommand\b{\beta}
\newcommand\g{\gamma}
\newcommand\sig{\sigma}

\newcommand\vareps{\varepsilon}
\newcommand\veps{\varepsilon}

\newcommand\e{\varepsilon}
\newcommand\estar{\varepsilon^*}
\newcommand\QLDP{{\rm QLDP}}
\newcommand\EA{{\rm EA}}
\newcommand\E{\mathcal{E}}
\renewcommand\O{\mathcal{O}}
\renewcommand\H{\mathcal{H}}

\newcommand\D{\mathcal{D}}
\newcommand\M{\mathcal{M}}
\providecommand\I{}
\renewcommand\I{\mathcal{I}}
\newcommand\tensor\otimes
\newcommand\ts\otimes
\DeclareMathOperator{\Tr}{Tr}

\newcommand\varempty{\varnothing} 

\newcommand\<\langle
\renewcommand\>\rangle
\renewcommand\v[1]{\boldsymbol{#1}}

\newtheorem{definition}{Definition}

\newtheorem{theorem}{Theorem}
\newtheorem{corollary}{Corollary}
\newtheoremstyle{italichead}{\topsep}{\topsep}{\itshape}{}{\itshape}{.}{.5em}{}
\theoremstyle{italichead}
\newtheorem{example}{Example}
\newtheorem{remark}{Remark}
\theoremstyle{plain}

\newenvironment{proofsketch}[1][Sketch of proof]{\par\noindent\textit{#1.}\ }{\par\smallskip}

\begin{document}
\begin{CJK*}{UTF8}{gbsn}
%
\title{Entanglement-Assisted Quantum Differential Privacy}

\author{
\IEEEauthorblockN{Chenjian Li$^{1,2,3}$, Ji Guan$^{1,2,*}$ and Mingsheng Ying$^4$}
\IEEEauthorblockA{$^1$Key Laboratory of System Software (Chinese Academy of Sciences), Beijing, China\\
$^2$Institute of Software, Chinese Academy of Sciences, Beijing, China\\
$^3$University of Chinese Academy of Sciences, Beijing, China\\
$^4$Centre for Quantum Software and Information,University of Technology Sydney, Sydney, Australia\\
Email: licj@ios.ac.cn, guanji1992@gmail.com, Mingsheng.Ying@uts.edu.au}
}


%


\maketitle

\begin{abstract}

Differential privacy has been successfully extended to quantum computing and information processing, where sensitive data are encoded in quantum states and protected by quantum noise mechanisms that keep the outcomes of any measurement on neighboring states nearly indistinguishable. In the quantum setting, however, the state submitted to a quantum computer may be entangled with a hidden reference system secretly held by an adversary. Such entanglement cannot be detected from the submitted state alone, and it poses a challenge with no counterpart in the classical setting. Existing frameworks of quantum differential privacy consider only measurements on the released state, leaving this scenario largely unexplored.

In this paper, we identify the \emph{entanglement attack}: an adversary submits one half of an entangled state to a quantum computer that runs a privacy mechanism, keeps the other half, and finally measures both halves jointly. We show that this attack may completely break a mechanism that satisfies quantum differential privacy with a finite parameter, driving its privacy parameter to infinity. To counter it, we propose \emph{entanglement-assisted quantum differential privacy} (EA-QDP), a privacy framework that is native to quantum entanglement and remains valid against such attacks. We characterize exactly when a mechanism has a finite EA-QDP parameter, and show that the optimal parameter $\varepsilon^*_{\mathrm{EA}}$ of any quantum mechanism can be computed by a semidefinite programming problem (SDP). Building on this, we quantify the extra privacy leakage caused by entanglement and show that it can be bounded by choosing appropriate mechanisms, with the depolarizing channel being our recommended choice among commonly used quantum noise mechanisms. Finally, we demonstrate both the attack and its countermeasure on simulated and physical quantum hardware, and the experimental results agree with our theoretical predictions.

\end{abstract}


%
\IEEEpeerreviewmaketitle

\section{Introduction}\label{sec:intro}
Quantum computing promises to significantly accelerate certain computational tasks by exploiting quantum phenomena such as superposition and entanglement. Landmark examples include Grover's algorithm for unstructured search~\cite{grover} and Shor's algorithm for integer factoring in polynomial time~\cite{shor}, a task for which no efficient classical algorithm is known. Beyond these algorithms, quantum computing is expected to bring substantial advances to areas such as chemistry and drug discovery~\cite{quantum_chem_review,quantum_computational_chem_review,quantum_drug_discovery_review}, financial analysis~\cite{quantum_finance_review}, and machine learning~\cite{quantum_ml_review}.
These prospects are steadily becoming reality as quantum hardware matures: quantum processors have demonstrated quantum computational advantage on different physical platforms~\cite{google_quantum_supremacy,zhu2022quantum,zhong2020quantum}, and quantum error correction has recently operated below the fault-tolerance threshold~\cite{google_qec_scaling,google_qec_below_threshold,bluvstein2024logical}, paving the way toward large-scale, fault-tolerant quantum computation.

Like their classical counterparts, however, quantum computers are exposed to security and privacy risks. Sensitive information enters a quantum computer in two forms. Some applications \emph{process inherently quantum data}, such as measurements from quantum sensor networks~\cite{spencer2026Differentially}, quantum states transmitted through communication channels~\cite{quantum_teleportation,chung2022design,sasaki2011field}, or datasets for quantum machine learning~\cite{quantum_ml_review}. Others \emph{process classical data quantumly}: Grover's search~\cite{grover} and quantum recommendation systems~\cite{qsve} encode sensitive classical data, such as medical histories or user preferences, into quantum states to achieve computational speedups. In both cases, the quantum computer runs operations that depend on private information. Various attacks on quantum systems have already been identified, such as coherent attacks~\cite{coherent_attack}, high-energy state attacks~\cite{xu2023securing}, and quantum side-channel attacks~\cite{xu2023exploration}, some of which exploit uniquely quantum properties to bypass classical defenses. Protecting private information on quantum computers therefore requires us to identify quantum-specific threats and to develop defenses against them in advance.

For classical data, differential privacy (DP)~\cite{dwork2013Algorithmic} has become the standard framework for protecting individual privacy. It injects random noise into the answers to queries on a database, so that the released output reveals little about any individual record. DP has been successfully applied to many areas, such as machine learning~\cite{abadi2016deep,ji2014differential}, recommendation systems~\cite{mcsherry2009Differentially,friedman2016Differential,chien2021Private,li2026Differential}, and quantum computing~\cite{zhou2017Differential,guan2025Optimal}. In particular, quantum differential privacy (QDP)~\cite{zhou2017Differential} extends DP to quantum data and protects sensitive quantum states. Later works studied QDP under different privacy and threat models, including R\'enyi QDP~\cite{hirche2023Quantum}, local quantum differential privacy~\cite{angrisani2025quantum,guan2025Optimal}, and QDP on product states, where two states are neighbors if one can be obtained from the other by a local operation on a single subsystem, mirroring the classical neighboring relation between databases~\cite{aaronson2019Gentle}.

In existing formulations of QDP, it is implicitly assumed that the released state is all the adversary gets: a state enters the quantum computer, is processed and perturbed by a noise mechanism, and the adversary measures only the released state. This is similar to the state distinguishing or tomography setting in physical experiments, where the entire state is controlled by the system manager. However, in the adversarial setting, a quantum adversary can prepare two \emph{entangled} systems, submit one of them to the quantum computer as an ordinary input, and secretly keep the other as hidden side information.

In classical DP, such side information would be harmless, since the guarantee holds regardless of what else the adversary knows. Quantum entanglement, however, is a form of correlation with no classical counterpart, and it is invisible to the quantum computer: the joint state carries information that neither system reveals on its own, and the submitted system alone looks just like an input that any honest user could have sent. This naturally raises several questions. \emph{Can the adversary take advantage of quantum entanglement and extract extra private information with the hidden system? If so, can it break the privacy guarantee completely? And how can we defend against such attacks?} Existing QDP frameworks analyze the released states without entanglement and cannot answer these questions.

In this paper, we answer all three questions. The answer to the first two is yes. We identify the \emph{entanglement attack} (Fig.~\ref{fig:threat-model}), in which the adversary entangles its input with a hidden system, submits the input as an ordinary user would, and measures the returned output together with the hidden system. Such a joint measurement can reveal far more than the output alone. Even against a mechanism that existing QDP frameworks certify to be private, a single pair of entangled qubits lets the adversary observe an outcome that lets it learn the secret with certainty, completely breaking the privacy guarantee.

For the third question, we propose \emph{entanglement-assisted quantum differential privacy} (EA-QDP), a privacy framework that accounts for the hidden system: its guarantee must hold even when the adversary measures the output together with any hidden system, which makes it strictly stronger than existing QDP notions. Such a guarantee may seem impossible to verify, since the hidden system can be arbitrarily large and there are infinitely many choices of reference systems; we show that it is nevertheless practical: a hidden system as large as the submitted primary system already captures all the leakage that entanglement can cause, and the privacy parameter of any mechanism can be computed by a semidefinite programming problem (SDP), a convex optimization problem that standard solvers handle efficiently. With this tool, we measure how much entanglement amplifies the leakage of a mechanism, and find that mechanisms that look equally private under existing frameworks can leak very differently under entanglement attacks. The depolarizing channel, which replaces the input with a completely random state with some probability, stands out: entanglement amplifies its leakage only mildly, by less than a factor of two on a single qubit, and it offers the best utility among all mechanisms with the same EA-QDP guarantee. Calibrated by EA-QDP, it thus provides a practical defense against entanglement attacks.

We validate these findings in noiseless simulation and on IBM quantum hardware, on a task with up to $10$ sensitive bits. Against a vulnerable mechanism calibrated by QDP, the entanglement attack recovers each bit with $100\%$ accuracy in simulation and $85.2\%$ on hardware, far beyond the $66.7\%$ that QDP permits. Once we switch to the depolarizing channel applied to each qubit and calibrated by EA-QDP, the attack stays within the permitted bound on hardware and simulation, and the bound is reached exactly in simulation, showing that the guarantee of EA-QDP is both sound and tight. In summary, our contributions are as follows:
\begin{enumerate}
  \item \textbf{Entanglement attack.} We identify the entanglement attack, an undetectable threat native to quantum systems, and show that it can completely break mechanisms that existing QDP frameworks deem private.
  \item \textbf{EA-QDP framework.} We propose EA-QDP, a privacy framework that remains valid under entanglement attacks, and show that its privacy parameter can be computed efficiently by an SDP. Using it, we quantify how much entanglement amplifies privacy leakage, and identify the depolarizing channel as a resistant mechanism with optimal utility.
  \item \textbf{Experimental validation.} We demonstrate the attack and the EA-QDP-based defense in simulation and on IBM quantum hardware, and the results agree with our theoretical predictions.
\end{enumerate}

The rest of the paper is organized as follows. Sec.~\ref{sec:preliminary} reviews the necessary background on quantum computing and differential privacy. Sec.~\ref{sec:threat-model} presents our threat model, and Sec.~\ref{sec:motivating-example} illustrates the entanglement attack with a motivating example. Sec.~\ref{sec:ea-qdp} introduces the EA-QDP framework, shows how to evaluate its privacy parameter, quantifies the extra leakage caused by entanglement and discusses potential defense strategies. Sec.~\ref{sec:exp} reports our experiments in simulation and on physical quantum hardware. Finally, Sec.~\ref{sec:related} discusses related work, and Sec.~\ref{sec:conclusion} concludes the paper.

\section{Preliminaries}\label{sec:preliminary}
In this section, we aim to provide readers with basic knowledge related to our work, including quantum computing and differential privacy. For a more comprehensive introduction, we refer the reader to~\cite{nielsen2000quantum} and~\cite{dwork2013Algorithmic} for quantum computation and differential privacy, respectively.
\subsection{Quantum Computing and Information }

\emph{Qubits and Dirac notation.} A \emph{quantum bit} (qubit) is described by a unit vector in $\mathbb C^2$, which is written in the Dirac notation as a \emph{ket}:
\begin{equation}
  |\psi\>=\a|0\>+\b|1\>,\qquad
  |0\>=\begin{pmatrix}1\\0\end{pmatrix},\quad
  |1\>=\begin{pmatrix}0\\1\end{pmatrix},
\end{equation}
where the complex amplitudes satisfy $|\a|^2+|\b|^2=1$. Unlike a classical bit, which takes the value either 0 or 1, a qubit can be in a \emph{superposition} of $|0\>$ and $|1\>$, such as
\begin{equation}
  |\pm\>=\frac{|0\>\pm|1\>}{\sqrt2}.
\end{equation}
The \emph{bra} $\<\psi|:=|\psi\>^\dagger$ is the conjugate transpose of $|\psi\>$, so that $\<\psi|\varphi\>$ denotes the inner product and $|\psi\>\<\varphi|$ the outer product of two vectors. More generally, the pure states of a $d$-level system are unit vectors in a Hilbert space $\H\cong\mathbb C^d$. The standard orthonormal basis $\{|x\>\}_{x=0}^{d-1}$ of $\H$ is called the \emph{computational basis}; for instance, $n$ qubits have $d=2^n$ and $2^n$ computational basis states $|00\cdots0\>,\ldots,|11\cdots1\>$.

\emph{Mixed states and density matrices.} A unit vector $|\psi\>$ describes a \emph{pure state}, which is known exactly. When a system is prepared in the pure state $|\psi_k\>$, with probability $p_k$ respectively, it is called a \emph{mixed state} and is described by the \emph{density matrix}
\begin{equation}
  \rho=\sum_k p_k|\psi_k\>\<\psi_k|,
\end{equation}
which is positive semidefinite ($\rho\ge0$) and has unit trace ($\Tr\rho=1$). We denote the set of all density matrices on $\H$ by $\D(\H)$. If every $|\psi_k\>$ is a computational basis state, $\rho=\sum_x p_x|x\>\<x|$ is diagonal and simply lists a probability distribution $\{p_x\}$; superpositions add off-diagonal entries, which have no classical counterpart.

Different preparations may lead to the same mixed state. Flipping a fair coin to prepare either $|0\>$ or $|1\>$, and flipping one to prepare either $|+\>$ or $|-\>$, both result in the \emph{maximally mixed state}
\begin{equation}
  \tfrac12|0\>\<0|+\tfrac12|1\>\<1|
  =\tfrac12|+\>\<+|+\tfrac12|-\>\<-|
  =\frac{I}{2}.
  \label{eq:max_mix_state}
\end{equation}
More generally, the maximally mixed state is $I/d$ on a $d$-level system. Since every prediction of quantum mechanics depends on $\rho$ alone, no physical test on the system can tell such preparations apart: the density matrix captures everything that can be learned from a system.

\emph{Composite systems and quantum entanglement.} When two systems $R$ and $S$ form a joint system, such combination can be described by the tensor product ``$\otimes $''. The Hilbert space for the joint system is $\H_R\ts\H_S$, whose basis states are $|r\>\ts|s\>$ that can usually be abbreviated as $|rs\>$. If $R$ and $S$ are prepared independently in $\rho_R$ and $\rho_S$, the joint state is the product $\rho_R\ts\rho_S$. Quantum mechanics, however, also allows joint states that admit no such decomposition. A pure joint state that cannot be written as a product $|\psi\>_R\ts|\varphi\>_S$ is said to exhibit \emph{quantum entanglement}, or simply to be \emph{entangled}. Entanglement has no classical counterpart, and it is the key resource behind many quantum protocols, as well as behind the attack studied in this paper. One of the best-known examples is the four \emph{Bell states}
\begin{equation}
  |\Phi^\pm\>=\frac{|00\>\pm|11\>}{\sqrt2},\qquad
  |\Psi^\pm\>=\frac{|01\>\pm|10\>}{\sqrt2},
\end{equation}
which form an orthonormal basis of the Hilbert space of two qubits.

For a composite system, its partial state on one system alone is obtained by the \emph{partial trace}:
\begin{equation}
  \rho_S=\Tr_R(\rho_{RS}):=\sum_r\big(\<r|\ts I_S\big)\,\rho_{RS}\,\big(|r\>\ts I_S\big),
\end{equation}
where $\{|r\>\}$ is any orthonormal basis of $\H_R$; the marginal $\rho_R=\Tr_S(\rho_{RS})$ is defined symmetrically. The partial trace sums out system $R$ along a basis, and describes everything that can be observed on $S$ alone. This is similar to calculating the marginal distribution of a joint probability distribution in probability theory. A direct calculation shows that all four Bell states have the same marginals $\rho_R=\rho_S=I/2$. Each half of a Bell pair thus looks exactly like a fair coin~\eqref{eq:max_mix_state} and reveals nothing about which Bell state the pair is in; yet the four Bell states are orthogonal, and hence perfectly distinguishable once both halves are examined together. The information of a Bell pair therefore resides entirely in the entanglement between its two halves, invisible to anyone who holds only one of them. This hidden entanglement is precisely what the attack in Sec.~\ref{sec:motivating-example} exploits.

\begin{table}[!t]
  \centering
  \caption{Concept Analogy between Classical and Quantum}
  \label{tab:dictionary}
  \small
  \begin{tabular}{@{}ll@{}}
    \toprule
    Classical & Quantum \\
    \midrule
    Bit $x\in\{0,1\}$ & Qubit $|\psi\>=\a|0\>+\b|1\>$ \\
    Probability distribution $\{p_x\}$ & Density matrix $\rho$ \\
    Joint distribution over $(R,S)$ & Joint state $\rho_{RS}$ \\
    Marginal distribution of $S$ & Partial trace $\Tr_R(\rho_{RS})$ \\
    (no counterpart) & Entanglement \\
    Deterministic computation & Unitary $\rho\mapsto U\rho U^\dagger$ \\
    Randomized mechanism & Noisy channel $\E$ \\
    Reading the output & Measurement $\{M_k\}$ \\
    \bottomrule
  \end{tabular}
\end{table}

\emph{Quantum channels.} Any physical process that transforms a quantum state is described by a \emph{quantum channel}. Every channel $\E$ admits a \emph{Kraus representation}~\cite{choi1975completely}
\begin{equation}
  \E(\rho)=\sum_\a K_\a\rho K_\a^\dagger,\qquad \sum_\a K_\a^\dagger K_\a=I,
\end{equation}
where the condition on the \emph{Kraus matrices} $\{K_\a\}$ ensures that $\E(\rho)$ is still a density matrix. A noiseless computation is the special case of a single unitary Kraus matrix $U$, namely $\rho\mapsto U\rho U^\dagger$. A noisy channel, such as those in Example~\ref{example:noise}, mixes several Kraus matrices and randomizes its output, which makes it the natural quantum analogue of a randomized mechanism in DP. We call $\E$ \emph{unital} if it maps the maximally mixed state to itself, i.e., $\E(I)=I$.

Two channels acting independently on two systems combine into their tensor product $\E\ts\mathcal F$, defined by $(\E\ts\mathcal F)(\rho_R\ts\rho_S)=\E(\rho_R)\ts\mathcal F(\rho_S)$ and extended linearly to all joint states. In particular, a channel $\E$ that touches only $S$ transforms a joint state $\rho_{RS}$ into $(\I_R\ts\E)(\rho_{RS})$, where $\I_R$ is the identity channel on $R$.

\emph{Measurements.} Classical information is extracted from a quantum system only through a \emph{measurement}. A measurement $\M=\{M_k\}_{k\in\O}$ with outcome set $\O$ is a collection of positive semidefinite matrices satisfying $\sum_k M_k=I$, known as a \emph{positive operator-valued measure} (POVM). On the state $\rho$, it yields the outcome $k$ with probability
\begin{equation}
  \Pr[\M(\rho)=k]=\Tr(M_k\rho).
\end{equation}
For instance, the computational-basis measurement $\{|0\>\<0|,|1\>\<1|\}$ yields $0$ and $1$ on $|\psi\>=\a|0\>+\b|1\>$ with probabilities $|\a|^2$ and $|\b|^2$, respectively. A more interesting example is, when the computational-basis measurement yields a uniformly random result on both $|+\>$ and $|-\>$ and cannot distinguish the two states, the measurement $\{|+\>\<+|,|-\>\<-|\}$ tells them apart with certainty. 

To summarize, Table~\ref{tab:dictionary} gathers the notions introduced above and compares them with their classical analogues. Readers familiar with probability but new to quantum computing may use it as a guide.

\subsection{Differential Privacy}
We now offer a brief introduction to classical and quantum differential privacy. For a general and more detailed introduction to differential privacy, we refer the reader to~\cite{dwork2013Algorithmic}.

The central idea of differential privacy (DP) is to release useful information about a population while limiting what the released output can reveal about any particular individual~\cite{li2025differential,spencer2026Differentially}. In particular, DP makes the queries probabilistic and requires the output distribution to remain nearly unchanged when one individual's data is modified. Consequently, an adversary is not able to reliably determine whether a particular record was used.

In the standard central model, a database $D=(x_1,\ldots,x_n)\in\mathcal{X}^n$ contains records from $n$ individuals and is held by a trusted curator, where $\mathcal {X}$ denotes the universe of all possible records, and $\mathcal Y$ is the set of possible results. The curator releases query results through a randomized mechanism $\M:\mathcal{X}^n\to\mathcal{Y}$. Two databases are called \emph{neighboring}, denoted by $D\sim D'$, if they differ in one record. The neighboring relation specifies the granularity of privacy, while the randomness of $\M$ dilutes the influence of each record on the released output. Formally, the differential privacy of a mechanism or database is defined as follows:

\begin{definition}[Differential privacy]
  A randomized mechanism $\M:\mathcal{X}^n\to\mathcal{Y}$ is $\vareps$-differentially private if, for every pair of neighboring databases $D\sim D'$ and every possible set of output events $E\subseteq\mathcal{Y}$,
  \begin{equation}
    \Pr[\M(D)\in E]\le e^{\vareps}\Pr[\M(D')\in E].
    \label{eq:classical-dp}
  \end{equation}
\end{definition}
Here, $\vareps$ bounds the multiplicative change between the two output distributions, and a smaller $\vareps$ provides stronger privacy. 
Next we briefly introduce the most basic mechanism, the randomized response.

\emph{Randomized response.} For a sensitive bit $x\in\{0,1\}$, randomized response reports its true value with probability $e^{\vareps}/(e^{\vareps}+1)$, and flips it before reporting with probability $1/(e^{\vareps}+1)$:
\begin{equation}
  \M_{\mathrm{RR}}(x)=
  \begin{cases}
    x,   & \text{with probability }\dfrac{e^{\vareps}}{e^{\vareps}+1},\\
    1-x, & \text{with probability }\dfrac{1}{e^{\vareps}+1}.
  \end{cases}
\end{equation}
For either reported value, the likelihood ratio between the two possible inputs is at most $e^{\vareps}$; hence, the mechanism satisfies $\vareps$-DP. Randomized response is also the prototype of \emph{local} differential privacy (LDP): each individual randomizes her own data before releasing it, so the guarantee must hold between \emph{any} two inputs of that individual, rather than between two databases that differ in one record.


In the quantum setting, classical data are replaced by quantum states, while the role of randomized mechanisms is played by quantum channels. A quantum mechanism $\E$ takes a state $\rho\in\D(\H)$ as input and produces an output state $\E(\rho)$, from which an adversary would try to extract classical information through quantum measurements. Consequently, quantum differential privacy requires the measurement outcome distributions to remain close when the input quantum states are neighboring.

\begin{definition}[QDP~\cite{zhou2017Differential} and QLDP~\cite{hirche2023Quantum}]
  A quantum mechanism $\E:\D(\H)\to\D(\H)$ is $\vareps$-quantum differentially private ($\vareps$-QDP) within $\eta\in[0,1]$ if, for any quantum states $\rho,\sig\in\D(\H)$ satisfying $T(\rho,\sig)\leq\eta$, any quantum measurement $\M=\{M_k\}_{k\in\mathcal{O}}$, and any set of outcomes $E\subseteq\mathcal{O}$,
  \begin{equation}
    \Pr[\M(\E(\rho))\in E]
    \leq e^{\vareps}\Pr[\M(\E(\sig))\in E],
    \label{eq:qdp}
  \end{equation}
  where $T(\rho,\sig):=\frac{1}{2}\Tr|\rho-\sig|\in[0,1]$ is the trace distance between $\rho$ and $\sig$, and $\Pr[\M(\E(\rho))\in E]=\sum_{k\in E}\Tr[M_k\E(\rho)]$.
  In particular, $\E$ is $\vareps$-quantum locally differentially private ($\vareps$-QLDP) if it is $\vareps$-QDP within $\eta=1$, that is, if~\eqref{eq:qdp} holds for every pair of quantum states.
\end{definition}
The threshold $\eta$ sets the neighboring relation between quantum states. QDP protects only input states that are close in trace distance, in analogy to the classical neighboring relation. QLDP follows the local model instead: the input state is the data of a single user, which the mechanism must conceal entirely, just as randomized response conceals the whole bit. Since the trace distance never exceeds one, $\eta=1$ makes every pair of states neighboring, which is exactly this requirement.

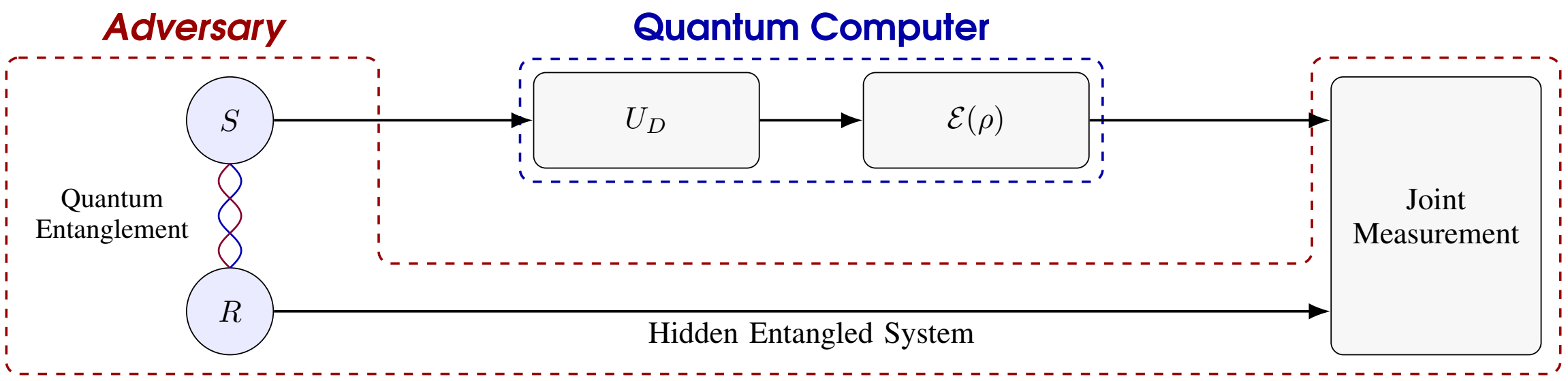
\begin{figure*}[!t]
  \centering
  \resizebox{0.85\textwidth}{!}{%
  \begin{tikzpicture}[
    font=\normalsize,
    box/.style={draw,rounded corners,minimum width=12mm,minimum height=8mm,align=center,fill=black!3},
    qcstep/.style={box,minimum width=26mm,minimum height=11mm},
    q/.style={draw,circle,minimum size=10mm,inner sep=0pt,fill=blue!8},
    arr/.style={-{Latex[length=2.5mm]},thick},
    advbox/.style={draw=red!60!black,dashed,thick,rounded corners,inner sep=2mm},
    qcbox/.style={draw=blue!65!black,dashed,thick,rounded corners,inner sep=1.5mm},
    head/.style={font=\large\fontfamily{pag}\fontseries{db}\fontshape{#1}\selectfont}
  ]
    \node[q] (S) at (0,0) {$S$};
    \node[q] (R) at (0,-2.2) {$R$};
    \draw[blue!70!black,line width=0.6pt,domain=0:1,samples=80,variable=\t]
      plot ({0.13*sin(540*\t)},{-0.5-1.2*\t});
    \draw[purple!70!black,line width=0.6pt,domain=0:1,samples=80,variable=\t]
      plot ({-0.13*sin(540*\t)},{-0.5-1.2*\t});
    \node[anchor=east,align=center,font=\small] (entlab) at (-0.35,-1.1) {Quantum\\Entanglement};
    \coordinate (advpad) at (1.5,-1.1);
    \node[advbox,draw=none,fit=(entlab)(S)(R)(advpad)] (advL) {};
    \node[qcstep] (U) at (4.8,0) {$U_D$};
    \node[qcstep] (E) at (8.6,0) {$\E(\rho)$};
    \node[qcbox,fit=(U)(E),
      label={[head=n,blue!65!black]above:Quantum Computer}] (qc) {};
    \node[box,minimum height=32mm,inner xsep=2.5mm] (meas) at (13.9,-1.1) {Joint\\Measurement};
    \node[advbox,draw=none,fit=(meas)] (advR) {};
    \coordinate (notch) at (0,-1.65);
    \draw[draw=red!60!black,dashed,thick,rounded corners]
      (advL.north west) -- (advL.north east) -- (advL.east |- notch)
      -- (advR.west |- notch) -- (advR.north west) -- (advR.north east)
      -- (advR.south east) -- (advL.south west) -- cycle;
    \node[head=sl,red!60!black,above] at (advL.north) {Adversary};
    \draw[arr] (S.east) -- (U.west);
    \draw[arr] (U) -- (E);
    \draw[arr] (E.east) -- (meas.west |- E.east);
    \draw[arr] (R.east) -- (meas.west |- R.east);
    \node[below] at (6.7,-2.2) {Hidden Entangled System};
  \end{tikzpicture}%
  }
  \caption{The threat model (quantum system and adversary) of Entanglement-Assisted Quantum Differential Privacy (EA-QDP). The adversary prepares an entangled state over $R\ts S$ but submits only system $S$ to the quantum computer, which runs a data-dependent computation $U_D$ followed by a privacy mechanism $\E$. System $R$ bypasses the quantum computer as a hidden entangled system. Finally, the adversary recombines the output with $R$ and extracts sensitive information through a joint measurement.}
  \label{fig:threat-model}
\end{figure*}

Next, we give some common channels that may serve as curating mechanisms for quantum differential privacy.
\begin{example}[Pauli noise mechanisms]\label{example:noise}
  A common way to erase the information in a qubit is to apply Pauli matrices
  \begin{equation}
    X=\begin{bmatrix}0&1\\1&0\end{bmatrix},\quad
    Y=\begin{bmatrix}0&-i\\i&0\end{bmatrix},\quad
    Z=\begin{bmatrix}1&0\\0&-1\end{bmatrix},
  \end{equation}
  which, together with the identity $I$, form a basis of $2\times2$ Hermitian matrices.
  \begin{enumerate}
    \item \emph{$X$-flip (bit-flip) mechanism:} The Pauli matrix $X$ flips a qubit in the computational basis, $X|x\>=|1-x\>$, just like a classical bit flip. Flipping the qubit with probability $1-p$ gives
    \begin{equation}
      \E_X(\rho)=p\rho+(1-p)X\rho X,
    \end{equation}
    with Kraus matrices $\{\sqrt{p}I,\sqrt{1-p}X\}$. For any $p$, it satisfies neither QDP within any $\eta>0$ nor QLDP. The Pauli matrices are cyclically  symmetric, so the $Y$ (bit-phase flip) and $Z$ (phase flip) noise mechanisms $\E_Y,\E_Z$ can be defined similarly with the same quantum differential privacy.
    \item \emph{Depolarizing mechanism:} The depolarizing mechanism replaces the input state with the maximally mixed state $I/2$ with probability $1-p$, and leaves it unchanged with probability $p$:
    \begin{equation}
      \mathcal D_{p}(\rho)=p\rho+(1-p)\frac{I}{2}.
    \end{equation}
    Equivalently, it applies each of $X$, $Y$, and $Z$ with probability $(1-p)/4$, which gives the Kraus representation
    \begin{align}
      \mathcal{D}_{p}(\rho)
      =\frac{1+3p}{4}\rho+\frac{1-p}{4}(X\rho X+Y\rho Y+Z\rho Z),
    \end{align}
    with Kraus matrices $\{\frac{\sqrt{1+3p}}{2}I, \frac{\sqrt{1-p}}{2}X,$ $\frac{\sqrt{1-p}}{2}Y,\frac{\sqrt{1-p}}{2}Z\}$. For $0\leq p<1$, it satisfies $\ln\!\left(1+\frac{2p\eta}{1-p}\right)$-QDP within $\eta$ and $\ln\!\left(\frac{1+p}{1-p}\right)$-QLDP.
  \end{enumerate}

\end{example}

\section{Quantum Threat Model}\label{sec:threat-model}

To analyze the privacy of a quantum system in the presence of external threats, we first need to characterize the behavior and capabilities of possible adversaries. In this section, we model the quantum system under attack and outline the threat model, highlighting two traits of the adversary: it can prepare entangled states, yet it is indistinguishable from an honest user. The illustration of the threat model is shown in Fig.~\ref{fig:threat-model}.

\emph{The quantum system.} Our threat model is based on a quantum computer that accepts quantum states as its inputs from users. On each input, it runs a computation $U_D$ to accelerate certain tasks that can involve sensitive data, such as whether a patient in a medical dataset has diabetes, or the preferences that a recommendation system holds about its users~\cite{qsve}. Before releasing the output state, the quantum computer applies a mechanism $\E$ that satisfies QDP, so that no measurement on the output can reveal much about any individual. Nothing seems wrong: a quantum state goes in, a noisy quantum state comes out, and the DP mechanism guarantees privacy of sensitive information.

\emph{Adversary's capability.} The adversary, however, does not have to submit a standalone state. It can prepare an entangled state on two systems, submit one of them to the quantum computer just as an ordinary user would, and secretly keep the other as a \emph{hidden entangled system}. Once the output returns, the adversary combines it with the hidden system and performs a joint measurement on both to extract sensitive information. We call this strategy an \emph{entanglement attack}. To enhance the robustness of our framework and the resilience of our defense strategies, we assume the adversary is free to choose the entangled state as well as the joint measurement. Without the hidden system, an entanglement attack reduces to the adversary of ordinary QDP~\cite{zhou2017Differential}, which may perform any measurement, but only on the released state and no extra entanglement is allowed.

Notably, entanglement attacks are invisible to the quantum computer.
From the quantum system's perspective, an entanglement attack needs nothing more than the ability to submit quantum states through the interface, which the quantum computer offers to every user anyway.
Whether a submitted state is entangled with a hidden system cannot be told from the submitted state alone: the partial state looks like an ordinary mixed state that an honest user could have prepared~\cite{hughston1993complete}.  Beyond what an ordinary user can do, the adversary only needs to prepare entangled states, be able to keep the hidden part until the output returns, and measure both parts jointly.


\section{Motivating Example: Entanglement Attack}\label{sec:motivating-example}
Before formalizing EA-QDP, we show that the threat in Sec.~\ref{sec:threat-model} is real. In this section, we build a toy quantum computer model whose privacy mechanism is as good as classical randomized response, and yet its privacy guarantee can be completely broken with an entanglement attack that uses a single Bell state. The example shows that a natural mechanism that provides a concrete QDP guarantee can fail catastrophically when the hidden reference system is allowed.

\emph{A mechanism that appears adequate.}
Let the sensitive data be a bit $x\in\{0,1\}$, and let the computation apply $U_D=X^x$ to the submitted qubit. In reality, the database and the algorithm may be much more sophisticated; for the sake of simplicity, here we use a simplified model to demonstrate the vulnerability. For the curating mechanism, a natural first attempt is the quantum version of randomized response:
\begin{equation}
  \E_{\mathrm{RR}}(\rho)=\frac12\big(\rho+X\rho X\big)=\E_X(\rho),
\end{equation}
which flips the data bit with probability $1/2$ and happens to be the bit-flip channel. Unfortunately, it leaves the superposed phase states $|+\>=\frac{|0\>+|1\>}{\sqrt2}$ and $|-\>=\frac{|0\>-|1\>}{\sqrt2}$ perfectly distinguishable, and does not satisfy \emph{quantum} differential privacy for any $\eta>0$. A possible way to fix it is to add a phase flip branch, resulting in the following quantum mechanism:
\begin{equation}
  \E_{XZ}(\rho)=\frac13\big(\rho+X\rho X+Z\rho Z\big).
\end{equation}
In the computational basis, this mechanism returns $x$ correctly with probability $2/3$ and flips the bit otherwise, still behaving like the randomized response mechanism; on the other hand, it now satisfies $\ln(1+\eta)$-QDP for $\eta\in[0,1]$.
Now the mechanism appears to be adequate to protect the privacy; what else can be wrong?

\emph{The entanglement attack.}
The twist is that the submitted qubit need not stand alone. Suppose the adversary prepares the Bell state $|\Phi^+\>$ on two qubits $R$ and $S$, keeps $R$ as the hidden entangled system, and submits only $S$, which returns after processing. Recall from Sec.~\ref{sec:preliminary} that each half of a Bell pair looks exactly like a fair coin $I/2$, while the four Bell states are perfectly distinguishable once both halves are examined together. The quantum computer thus receives a qubit that any honest user might have sent, whereas the adversary can treat the four Bell states as four \emph{labels} of the pair $(R,S)$: a joint measurement in the Bell basis, called the \emph{Bell measurement}, can read the label without error.

After the quantum computer returns $S$, the adversary performs the Bell measurement on the pair $(R,S)$. A direct calculation gives the probability of observing each label, as summarized in Table~\ref{tab:motivating}. Before the mechanism, the pair is in the Bell state $|\Phi^+\>$ if $x=0$ and in $|\Psi^+\>$ if $x=1$, so the Bell measurement would reveal $x$ exactly and a curating mechanism is needed. The mechanism $\E_{XZ}$ spreads each secret evenly over three of the four labels, yet, unfortunately, for each secret the remaining label still has probability $0$. In particular, examine the Table~\ref{tab:motivating} and we (and the adversary) can see that,
\begin{equation}
  \Pr[\Phi^-|x=0]=\tfrac13,\qquad \Pr[\Phi^-|x=1]=0.
  \label{eq:motivating-zero-event}
\end{equation}

\begin{table}[!t]
  \centering
  \caption{Outcome probabilities of the Bell measurement in the motivating example}
  \label{tab:motivating}
  \small
  \setlength{\tabcolsep}{3.5pt}
  \newcommand{\leak}{\textcolor{red!80!black}{$\mathbf{0}$}}
  \begin{tabular}{@{}c|cccc|cccc@{}}
    \toprule
    & \multicolumn{4}{c|}{After $U_D$ applies} & \multicolumn{4}{c}{After $\E_{XZ}$ applies} \\
    \cmidrule(lr){2-5}\cmidrule(lr){6-9}
    Secret & $\Phi^+$ & $\Phi^-$ & $\Psi^+$ & $\Psi^-$ & $\Phi^+$ & $\Phi^-$ & $\Psi^+$ & $\Psi^-$ \\
    \midrule
    $x=0$ & $1$ & $0$ & $0$ & $0$ & $1/3$ & $1/3$ & $1/3$ & \leak \\
    $x=1$ & $0$ & $0$ & $1$ & $0$ & $1/3$ & \leak & $1/3$ & $1/3$ \\
    \bottomrule
  \end{tabular}
  \vspace{-.4em} 
\end{table}

Consequently, observing the outcome $\Phi^-$ allows the adversary to conclude with certainty that the secret bit is $0$. If the bit encodes a sensitive attribute, such as whether a patient has diabetes, a single occurrence of this outcome exposes the patient's privacy without mercy. This is exactly what differential privacy was designed to prevent: a mechanism that satisfies DP should prevent an adversary from learning much about the sensitive data, \emph{no matter what the outcome is.} In the language of DP, since no finite multiplicative factor can bound $(1/3)/0$, the finite $\ln(1+\eta)$-QDP guarantee of $\E_{XZ}$ collapses to $+\infty$ under the entanglement-assisted threat model. This single zero-probability event motivates the stronger privacy notion introduced next.

\section{Entanglement-Assisted QDP}\label{sec:ea-qdp}

In this section, we formally introduce \emph{entanglement-assisted quantum differential privacy} (EA-QDP), a privacy framework that accounts for adversaries holding a hidden entangled system. We begin with the definition of EA-QDP and examine its relation to the existing notions of quantum differential privacy, then  show that the adversary never needs a reference system larger than the system it submits. Based on the dimension bound of effective reference systems, we finally study how to evaluate the $\vareps$ parameter of a quantum channel under EA-QDP, and reveal that the evaluation of a quantum channel is equivalent to an semidefinite programming (SDP) problem. To streamline the reading experience, we defer all proofs to the appendix and provide proof sketches in the main text when possible.

In accordance with the threat model in Sec.~\ref{sec:threat-model}, the adversary may entangle the submitted primary system $S$ with a reference system $R$, and measure both systems jointly after the mechanism. To safeguard privacy under such conditions, we define EA-QDP as follows.

\begin{definition}[Entanglement-Assisted Quantum Differential Privacy]\label{def:ea-qdp}
  A quantum mechanism $\E:\D(\H_S)\to\D(\H_S)$ is $\vareps$-\emph{entanglement-assisted quantum differentially private} ($\vareps$-EA-QDP) if,
  \begin{enumerate}
    \item for any finite-dimensional reference system $R$,
    \item for any quantum states $\rho_{RS},\sig_{RS}\in\D(\H_R\ts\H_S)$ that have identical marginal states on the reference system $R$, i.e., $\rho_R=\sig_R,$ where $\rho_R:=\Tr_S(\rho_{RS}),\sig_R:=\Tr_S(\sigma_{RS}),$
    \item and for any quantum measurement $\M=\{M_k\}_{k\in\O}$ on $R\ts S$ and any set of measurement outcomes $E\subseteq\O$,


  \end{enumerate}
   the following inequality holds:
  \begin{multline}
    \Pr\!\big[\M\big((\I_R\ts\E)(\rho_{RS})\big)\in E\big] \\
    \le e^{\vareps}\Pr\!\big[\M\big((\I_R\ts\E)(\sig_{RS})\big)\in E\big].
    \label{eq:ea-qdp}
  \end{multline}
\end{definition}

The definition mirrors the structure of our threat model elaborated  in Sec.~\ref{sec:threat-model} and Sec.~\ref{sec:motivating-example}: even if the adversary holds the hidden entangled system and measures it together with the released system, it should not be able to break the privacy of an $\vareps$-EA-QDP channel.
In the definition, $\rho_S,\sigma_S$ correspond to the quantum states after the quantum computer's computation process $U_D$, and thus the sensitive data may have been attached to them; on the other hand, the marginal states $\rho_R=\sigma_R$ correspond to the hidden entangled system held by the adversary and are therefore independent of sensitive data. Since we make no assumption on what the quantum computer computes, any marginal states $\rho_S, \sigma_S$ are allowed on $S$, enhancing the robustness and generality of our framework.
As the reference system is not present to the quantum computer, the curating mechanism acts on $S$ only, resulting in a total channel $\I_R\ts\E$ on the joint system $R\otimes S$---the identity channel $\I_R$ expresses that $R$ is bypassed by the quantum computer.



Having defined EA-QDP, we now compare it with the existing notions of quantum differential privacy that do not involve entanglement. Clearly, EA-QDP is an enhancement of QDP: by taking $R$ to be a one-dimensional system, the constraint $\rho_R=\sig_R$ holds trivially, and as we make no restriction on $\rho_S,\sigma_S$, Definition~\ref{def:ea-qdp} reduces exactly to QDP within $\eta=1$, i.e., QLDP, which in turn implies QDP within any $\eta$. The converse does not hold, as the motivating example shows: $\E_{XZ}$ satisfies $\ln 2$-QLDP, yet has no finite EA-QDP parameter. EA-QDP is thus strictly stronger than QDP. We summarize this hierarchy as follows:

\begin{theorem}[EA-QDP $\Rightarrow$ QDP within any $\eta$]\label{lemma:hierarchy}
  If a quantum mechanism $\E$ is $\vareps$-EA-QDP, then it is $\vareps$-QDP within any $\eta\in[0,1]$; in particular, it is also $\vareps$-QLDP.
\end{theorem}

Definition~\ref{def:ea-qdp} gives an operational definition of EA-QDP, but it is hard to verify directly, since it quantifies over all reference systems, which may be infinitely large. Fortunately, as we show next, it suffices to check reference systems no larger than the primary system.

\begin{theorem}[Dimension bound on the reference system]\label{thm:dim-bound}
  Let $d=\dim(\H_S)$. A quantum mechanism $\E$ is $\vareps$-EA-QDP if and only if~\eqref{eq:ea-qdp} holds for all reference systems $R$ with $\dim(\H_R)\le d$.
\end{theorem}

\begin{proofsketch}
  Suppose that some pair of states $\rho_{RS},\sig_{RS}$ with $\rho_R=\sig_R$ and some measurement violate~\eqref{eq:ea-qdp} with a large reference system $R$; we call these states the \emph{witness states} of the privacy violation.
  The Schmidt decomposition of bipartite states shows that any such states can be \emph{embedded} into $r\ts S$ for a smaller reference system $r$ with $\dim(\H_r)\le d$, so the embedded states still violate~\eqref{eq:ea-qdp} on the smaller reference system. Every violation of $\vareps$-EA-QDP can therefore be found on a reference system of dimension at most $d$, and it suffices to check such dimension-bounded reference systems.
\end{proofsketch}

Theorem~\ref{thm:dim-bound} replaces the requirement of checking over infinitely many reference systems in Definition~\ref{def:ea-qdp} with a much easier one, and makes the evaluation of the $\vareps$ parameter possible under the EA-QDP framework. Since any smaller reference system can also be embedded into a larger one, it actually suffices to check a \emph{single} reference system that has exactly the same dimension as $S$. For the adversary, this means that a hidden entangled system as large as the submitted one already achieves all the leakage that entanglement can bring; the adversary gains no benefits from a reference system larger than the primary system. 


\subsection{Evaluating EA-QDP}\label{sec:evaluating-ea-qdp}

In this section, we discuss the evaluation of the parameter $\veps$ for EA-QDP, providing a foundation for quantitative analysis of mechanisms within EA-QDP.
Our main result shows that this can be computed exactly by a semidefinite programming (SDP) problem, whose only input is the joint state that the adversary holds after the entanglement attack. Along the way, we also obtain a simple test for when a mechanism is completely broken, and closed-form values for the noise mechanisms considered so far.

As in classical DP, a mechanism that is $\vareps$-EA-QDP is also $\vareps'$-EA-QDP for every $\vareps'\ge\vareps$. What characterizes a mechanism is therefore the smallest admissible $\vareps$, which we define as follows.

\begin{definition}[Optimal EA-QDP Parameter]\label{def:ea-qdp-value}
  The optimal EA-QDP parameter of a quantum mechanism $\E$ is
  \begin{equation}
    \vareps^*_{\rm EA}(\E):=\inf\{\vareps\ge0:\ \E\text{ is }\vareps\text{-EA-QDP}\}.
    \label{eq:ea-qdp-param}
  \end{equation}
  If $\E$ is not $\vareps$-EA-QDP for any finite $\vareps$, we write $\vareps^*_{\rm EA}(\E)=+\infty$ or say that $\E$ does not satisfy EA-QDP.
\end{definition}

Operationally, $\vareps^*_{\rm EA}(\E)$ is the largest log-likelihood ratio that an entanglement-assisted adversary can ever observe between two inputs with the same marginal on $R$, whatever entangled states and joint measurements it uses. The optimal QDP parameter $\vareps^*_{\rm QDP}(\E)$ (within $\eta$) and the optimal QLDP parameter $\vareps^*_{\rm QLDP}(\E)$ are defined in the same way, so that the three can be compared directly. The mechanism $\E_{XZ}$ in Sec.~\ref{sec:motivating-example} shows they can diverge significantly: it has $\vareps^*_{\rm QDP}(\E_{XZ})=\ln(1+\eta)$, yet $\vareps^*_{\rm EA}(\E_{XZ})=+\infty$.

To evaluate $\vareps^*_{\rm EA}$, a key quantity we will be focusing on is
\begin{equation}
  J_\E:=(\I_R\ts\E)(|\Omega\>\<\Omega|)
  \label{eq:choi}
\end{equation}
where $|\Omega\>:=\sum_{i=1}^d|i\>_R|i\>_S$ is the (unnormalized) maximally entangled state between two $d$-dimensional systems. This matrix models the joint output state in the entanglement attack threat model: suppose the adversary is attacking a $d$-dimensional system with the maximally entangled state $|\Omega\>$ (up to a normalization), submits $S$ to the quantum computer, and keeps $R$. Once the mechanism has acted, the adversary then derives the joint state $(\I_R\ts\E)(|\Omega\>\<\Omega|)$, up to the factor $1/d$ and the unitary computation. $J_{\E}$ therefore captures the essence of the entanglement attack.
Notably, in quantum information theory, $J_\E$ is also known as the \emph{Choi--Jamio\l{}kowski matrix} of $\E$. It is able to completely determine $\E$, and characterizes its behavior upon entangled states. As we show next, it also determines how private $\E$ is against entanglement attacks.

\emph{Classical intuition.} Before we proceed, it helps to look at the classical analogy first. Suppose a classical local DP mechanism takes an input $x$ and outputs $y$ with probability $p(y|x)$; it is $\vareps$-LDP if $p(y|x)\le e^{\vareps}p(y|x')$ for all $x,x',y$. This condition can be rewritten by inserting an \emph{input-independent} function in between: the mechanism is $\vareps$-LDP if and only if there is a nonnegative function $q(y)$, for example $q(y)=\max_x p(y|x)$, such that
\begin{equation}
  p(y|x)\le q(y)\le e^{\vareps}p(y|x'),\qquad\forall\,x,x',y.\label{eq:ldp-sandwich}
\end{equation}
Since $q$ does not depend on the input, it describes, up to normalization, a \emph{perfectly private} mechanism that reveals nothing about $x$. In other words, a mechanism is $\vareps$-LDP exactly when its output distribution stays within a factor of $e^{\vareps}$ of a perfectly private mechanism.

The same picture carries over to the quantum case. The quantum counterpart of a perfectly private mechanism is the \emph{replacement map} $\mathcal R_\omega(\rho):=\Tr(\rho)\,\omega$, which completely discards its input and always outputs the same operator $\omega\ge0$. Even an entanglement-assisted adversary learns nothing from it, since the joint output $(\I_R\ts\mathcal R_\omega)(\rho_{RS})=\rho_R\ts \omega$ breaks the relation between $\omega$ and the $\rho_R$ marginal that the adversary holds. In particular, the Choi matrix of the replacement map reads $J_{\mathcal R_\omega}=I_R\ts \omega$. Our main result shows that the classical sandwich~\eqref{eq:ldp-sandwich} generalizes to EA-QDP in the quantum context, resulting in an inequality over matrices instead. Here, for Hermitian matrices, $A\le B$ means that $B-A$ is positive semidefinite.

\begin{theorem}[Evaluating EA-QDP by SDP]\label{thm:cp-sandwich}
  A quantum mechanism $\E$ is $\vareps$-EA-QDP if and only if there is a Hermitian operator $\omega\ge0$ such that
  \begin{equation}
    J_\E\le I_R\ts \omega\le e^{\vareps}J_\E.
    \label{eq:cp-sandwich}
  \end{equation}
  Consequently, $\vareps^*_{\rm EA}(\E)=\ln\gamma^*$, where $\gamma^*$ is the optimal value of the SDP problem
  \begin{align}
    \mathrm{minimize}\quad & \gamma\nonumber\\
    \mathrm{over}\quad & \gamma\in\mathbb R,\quad \omega\ge0 \text{ on } \H_S,\nonumber\\
    \mathrm{subject\ to}\quad & J_{\E}\le I_R\ts \omega\le\gamma J_{\E},
    \label{eq:sdp}
  \end{align}
  and $\gamma^*=+\infty$ if~\eqref{eq:sdp} is infeasible.
\end{theorem}

Theorem~\ref{thm:cp-sandwich} shows that a mechanism is $\vareps$-EA-QDP if it is close to a perfectly private quantum mechanism, a quantum analogue of the classical relation~\eqref{eq:ldp-sandwich}. Technically, the entanglement is characterized and captured by the Choi matrix and the maximally entangled state $|\Omega\>$.
Sufficiency of~\eqref{eq:cp-sandwich} is straightforward: for any two inputs with $\rho_R=\sig_R$,
\begin{equation}
  (\I_R\ts\E)(\rho_{RS})\le\rho_R\ts \omega=\sig_R\ts \omega\le e^{\vareps}(\I_R\ts\E)(\sig_{RS}),
\end{equation}
so that no joint measurement can tell the two outputs apart by more than a factor of $e^{\vareps}$. The converse direction is more involved, and we defer its technical proof to Appendix~\ref{app:proof-cp-sandwich}.

Theorem~\ref{thm:cp-sandwich} replaces the definition of EA-QDP, which ranges over all reference systems, input states, and joint measurements, with a single matrix inequality. The SDP formulation then makes EA-QDP easy to evaluate in practice. For QDP and QLDP, the privacy parameter is evaluated by maximizing a non-convex objective over measurements or pure states~\cite[Theorem~4.1]{guan2023Detectinga},~\cite[Theorem~2]{guan2025Optimal}. The SDP~\eqref{eq:sdp}, in contrast, is a convex optimization problem which can be solved by standard solvers, and its optimum is certified. In terms of complexity, the SDP problem~\eqref{eq:sdp} has $d^2+1$ real variables and two linear matrix inequalities of size $d^2\times d^2$, so standard interior-point methods solve it in time $O(d^9)$~\cite{boyd2004convex}. In practice, numerically evaluating $\estar_\EA$ via SDP typically scales to dimensions around $d=10$ for arbitrary random channels, and to around $d=35$ for highly symmetric channels such as the depolarizing channel.

\emph{When does entanglement break a mechanism completely?} The SDP gives the exact value of $\vareps^*_{\rm EA}$, but whether this value is finite can be decided even more directly, from the positivity of $J_\E$.

\begin{theorem}[Finiteness of $\vareps^*_{\rm EA}$]\label{thm:ea-finite}
  Let $\E$ be a quantum mechanism with $\E(I)>0$ (e.g., any unital mechanism). Then $\vareps^*_{\rm EA}(\E)<+\infty$ if and only if $J_\E>0$, i.e., $J_\E$ has no zero eigenvalue.
\end{theorem}

Theorem~\ref{thm:ea-finite} has a direct operational interpretation. A zero eigenvalue of $J_\E$ means that, under the entanglement attack, some joint measurement outcome never occurs. If the outcome occurs for some other secret value, then the adversary immediately learns with certainty that the secret takes the other value. Table~\ref{tab:motivating} in Sec.~\ref{sec:motivating-example} is a direct example of this phenomenon. 

\begin{remark}[An erratum on~\cite{guan2025Optimal}]\label{remark:erratum}
  When comparing our result with prior work on QLDP, we note an error in Theorem~3 of~\cite{guan2025Optimal}. For mechanisms with $\E(I)>0$, Theorem~3 of~\cite{guan2025Optimal} states that $\vareps^*_{\rm QLDP}(\E)<+\infty$ if and only if $(\E^\dagger\ts\I)(|\Omega\>\<\Omega|)>0$, where $\E^\dagger$ is the adjoint of $\E$; this condition is equivalent to $J_\E>0$. By Theorem~\ref{thm:ea-finite}, the condition in fact characterizes $\vareps^*_{\rm EA}(\E)<+\infty$. It is therefore sufficient for a finite QLDP parameter, but not necessary. The correct criterion for QLDP is that $\E(|\psi\>\<\psi|)>0$ for every pure state $|\psi\>$. For the same reason, Corollary~2 of~\cite{guan2025Optimal}, which states that a mechanism with fewer than $d^2$ Kraus matrices has $\vareps^*_{\rm QLDP}(\E)=+\infty$, holds for EA-QDP rather than for QLDP.
\end{remark}

Finally, we apply the SDP to single-qubit Pauli mechanisms, which include all mechanisms in Example~\ref{example:noise} and Sec.~\ref{sec:motivating-example}. For these mechanisms, the SDP can be solved in closed form.

\begin{theorem}[EA-QDP of Pauli mechanisms]\label{thm:ea-values}
  For a single-qubit Pauli mechanism
  \begin{equation}
    \E(\rho)=\sum_{\alpha\in\{I,X,Y,Z\}}w_\alpha\,\sigma_\alpha\rho\sigma_\alpha,
  \end{equation}
  where $(\sigma_I,\sigma_X,\sigma_Y,\sigma_Z)=(I,X,Y,Z)$, $w_\alpha\ge0$, and $\sum_\alpha w_\alpha=1$, we have
  \begin{equation}
    e^{\vareps^*_{\rm EA}(\E)}=\frac{\max_\alpha w_\alpha}{\min_\alpha w_\alpha},
  \end{equation}
  and $\vareps^*_{\rm EA}(\E)=+\infty$ if some $w_\alpha=0$. In particular, $\vareps^*_{\rm EA}=+\infty$ for $\E_X$, $\E_Y$, $\E_Z$ (for every $p$) and for $\E_{XZ}$.

  It is not until we collect all three types of Pauli errors that we obtain the depolarizing noise that has finite $\estar_{\EA}$ and is resistant against entanglement attacks. For the depolarizing channel $\D_p$, we have:
    \begin{equation}
      \estar_\EA(\D_p)=\left\{\begin{array}{ll}
        \ln\frac{1+3p}{1-p} & \text{ for } 0\le p<1,\\
        +\infty & \text{ for } p=1.
      \end{array}\right.
    \end{equation}
\end{theorem}

\subsection{Defense Strategy: Entanglement Amplification}\label{sec:amplification}
With the privacy parameters now computable, we return to the central question of this paper: how much additional privacy leakage does entanglement cause? In this section we answer this question quantitatively and give suggestions on how to select mechanisms that are more resistant against entanglement attacks. Recall from Theorem~\ref{lemma:hierarchy} that $\e$-EA-QDP implies $\e$-QDP and $\e$-QLDP for a channel; rephrasing the relation quantitatively immediately yields the following inequality:

\begin{corollary}[Hierarchy of privacy parameters]\label{cor:hierarchy}
  For any quantum mechanism $\E$ and any $\eta\in[0,1]$,
  \begin{equation}
    \vareps^*_{\rm EA}(\E)\ge\vareps^*_{\rm QLDP}(\E)\ge\vareps^*_{\rm QDP}(\E).
    \label{eq:hierarchy}
  \end{equation}
\end{corollary}


In the hierarchy, the QDP privacy model with $\eta=1$, i.e., QLDP, is the strongest DP requirement for a quantum mechanism without entanglement. For a mechanism $\E$, a nonzero gap $\estar_\EA(\E)-\estar_\QLDP(\E)>0$ therefore quantifies the additional privacy leakage caused by entanglement alone, i.e., the extra information an adversary can extract through an entanglement attack. Hence, we can further define the \emph{entanglement amplification} of a channel that captures this leakage.

\begin{definition}[Entanglement Amplification]
  For a quantum mechanism $\E:\D(\H_S)\to\D(\H_S)$ with $\vareps^*_{\rm QLDP}(\E)<+\infty$, its multiplicative entanglement amplification is defined as
  \begin{equation}
    A_{\rm ent}(\E):=\frac{\exp(\vareps_{\rm EA}^*(\E))}{\exp(\vareps_{\rm QLDP}^*(\E))}\ge 1.
  \end{equation}
  Or equivalently, its entanglement amplification is defined as:
  \begin{equation}
    \Delta_{\rm ent}(\E):=\ln\big(A_{\rm ent}(\E)\big)=\vareps_{\rm EA}^*(\E)-\vareps_{\rm QLDP}^*(\E)\ge 0.
  \end{equation}
\end{definition}

Intuitively, $A_{\rm ent}$ measures by how much the hidden entangled system contributes to the likelihood ratio that the adversary can observe between two neighboring inputs, and hence how much more it can learn about the sensitive data. A larger $\Delta_{\rm ent}(\E)$ means that a mechanism is more susceptible to entanglement attacks, and a smaller one means that it is more resistant. For example, the channel $\E_{XZ}$ in the motivating example has
\begin{align}
  \Delta_{\rm ent}(\E_{XZ})&=\estar_\EA(\E_{XZ})-\estar_\QLDP(\E_{XZ})\nonumber\\&=+\infty-\ln 2=+\infty,
\end{align}
which means its privacy guarantee is completely broken by the entanglement attack of Sec.~\ref{sec:motivating-example}, although it satisfies $\ln(1+\eta)$-QDP when entanglement is not allowed. To protect sensitive data against such attacks, a data curator should at least choose a mechanism whose entanglement amplification is bounded, preferably small. In the rest of this section, we analyze the entanglement amplification of several representative classes of mechanisms, revealing which kinds of channels are less susceptible to entanglement attacks.

\emph{Quantum-to-classical channels: fully resistant, but useless for quantum computing.} The most resistant mechanisms are the mechanisms that measure the input and release only a classical outcome:
\begin{equation}
  \E_{\text{Q-to-C}}(\rho)=\sum_y\Tr(M_y\rho)\,|y\>\<y|,
\end{equation}
where $\{M_y\}$ is a measurement with $M_y>0$ for every $y$, and $\{|y\>\}$ is an orthonormal basis. For such mechanisms,
\begin{equation}
  e^{\estar_\EA(\E_{\text{Q-to-C}})}=e^{\estar_\QLDP(\E_{\text{Q-to-C}})}=\max_y\frac{\lambda_{\max}(M_y)}{\lambda_{\min}(M_y)},
\end{equation}
so that $A_{\rm ent}=1$ and $\Delta_{\rm ent}=0$: entanglement brings no extra leakage at all. For quantum computing, however, the price is prohibitive. The mechanism destroys all quantum coherence of its output, so any subsequent quantum processing, and with it any quantum speedup, is lost. Quantum-to-classical channels are therefore not an option for quantum computers in general.

\emph{Near-replacement channels and the depolarizing channel: resistant and useful.} Mechanisms that preserve coherence generally cannot reach $A_{\rm ent}=1$, but they can keep it small. Recall from Sec.~\ref{sec:evaluating-ea-qdp} that a replacement map $\mathcal R_\omega$, which outputs a fixed state $\omega$, is perfectly private. A mechanism whose $J_\E$ stays close to $J_{\mathcal R_\omega}=I_R\ts\omega$ inherits a bounded privacy parameter: if
\begin{equation}
  (1-\alpha)\,I_R\ts\omega\le J_\E\le(1+\alpha)\,I_R\ts\omega
\end{equation}
for some $0\le\alpha<1$, then Theorem~\ref{thm:cp-sandwich} with $(1+\alpha)\omega$ replacing $\omega$ gives
\begin{equation}
  A_{\rm ent}(\E)\le e^{\estar_\EA(\E)}\le\frac{1+\alpha}{1-\alpha}.
\end{equation}

The depolarizing channel $\D_p(\rho)=p\rho+(1-p)I/d$ on a $d$-dimensional system is a special case of this class. It replaces the input state with the maximally mixed state with probability $1-p$, and by symmetry it treats all directions alike. Its parameters are
\begin{equation}
  e^{\estar_\QLDP(\D_p)}=1+\frac{dp}{1-p},\quad e^{\estar_\EA(\D_p)}=1+\frac{d^2p}{1-p},\label{eq:dep-values}
\end{equation}
so that
\begin{equation}
  A_{\rm ent}(\D_p)=\frac{1-p+d^2p}{1-p+dp}<d.
\end{equation}
For a qubit with $d=2$, $A_{\rm ent}(\D_p)=(1+3p)/(1+p)<2$: the hidden entangled system can never double the likelihood ratio. For instance, $p=0.5$ gives $e^{\estar_\QLDP}=3$, $e^{\estar_\EA}=5$, and $A_{\rm ent}\approx1.67$, while even the weak noise $p=0.9$ gives $e^{\estar_\QLDP}=19$, $e^{\estar_\EA}=37$, and $A_{\rm ent}\approx1.95$. Other near-replacement channels enjoy similar bounds, but in terms of utility the depolarizing channel is the best choice:  among all unital mechanisms with a given QLDP parameter, it maximizes both the fidelity and the trace-distance utility of the output~\cite{guan2025Optimal}. In fact, the optimal utility property of $\D_p$ still remains when we fix the EA-QDP parameter instead:

\begin{theorem}[Optimal utility under EA-QDP]\label{thm:optimal-utility}
  For a quantum mechanism $\E$ on a $d$-dimensional system, define its fidelity utility and anti-trace distance utility as the worst-case closeness between an input pure state and its output~\cite{guan2025Optimal}:
  \begin{equation}
    \begin{aligned}
      F(\E)&:=\min_{\psi}\<\psi|\E(\psi)|\psi\>,\\
      \hat T(\E)&:=1-\max_{\psi}T\big(\E(\psi),\psi\big),
    \end{aligned}
  \end{equation}
  where $\psi:=|\psi\>\<\psi|$ ranges over all pure states. If $\E$ is $\vareps$-EA-QDP, then
  \begin{equation}
    F(\E),\ \hat T(\E)\le\frac{e^{\vareps}+d-1}{e^{\vareps}+d^2-1},
    \label{eq:optimal-utility}
  \end{equation}
  for any channel, and both bounds are attained by the depolarizing channel $\D_{p^*}$ with $p^*=(e^{\vareps}-1)/(e^{\vareps}+d^2-1)$.
\end{theorem}

Combining optimal utility and a small and bounded $A_{\rm ent}$, the depolarizing channel is therefore our recommended mechanism against entanglement attacks.

\emph{Non-uniform Pauli channels: vulnerable.} Once the noise becomes anisotropic, the entanglement amplification can grow without bound. Consider a Pauli mechanism
\begin{equation}
  \E(\rho)=\sum_{\alpha\in\{I,X,Y,Z\}}w_\alpha\,\sigma_\alpha\rho\sigma_\alpha,\qquad w_\alpha>0.
\end{equation}
By~\cite{guan2025Optimal} and Theorem~\ref{thm:ea-values},
\begin{equation}
  e^{\estar_\QLDP(\E)}=\frac{1+s}{1-s},\quad e^{\estar_\EA(\E)}=\frac{\max_\alpha w_\alpha}{\min_\alpha w_\alpha},\label{eq:two_fml}
\end{equation}
where $s$ is the largest factor by which $\E$ shrinks the Bloch vector of a single qubit along the $X$, $Y$, or $Z$ axis, which can be computed as
\begin{equation}
  \begin{aligned}
    s_X&=w_I+w_X-w_Y-w_Z,\\
    s_Y&=w_I-w_X+w_Y-w_Z,\\
    s_Z&=w_I-w_X-w_Y+w_Z,
  \end{aligned}
  \quad
  s:=\max_{\alpha\in\{X,Y,Z\}}|s_\alpha|.
\end{equation}
The two formulas in~\eqref{eq:two_fml} reflect what each adversary can observe, which are quite different. Without entanglement, the adversary only sees the net effect of the noise on a single qubit: the four Pauli errors blur the qubit together, and their weights enter only through the combined contraction $s$. With a hidden entangled qubit, the picture changes completely. If the adversary submits one half of the Bell state $|\Phi^+\>$, each of the four Pauli errors turns the joint state into a different Bell state, and a joint measurement distinguishes the four Bell states perfectly. The adversary can thus tell exactly which error has occurred, and the rarest error becomes the most revealing: it produces an outcome that the mechanism almost never outputs, just as $\E_{XZ}$, which lacks the $Y$ error, never outputs $|\Psi^-\>$ in Sec.~\ref{sec:motivating-example}.
To see the effect, take the weights $(w_I,w_X,w_Y,w_Z)=(0.5,0.3,0.1,0.1)$ and $(0.5,0.3,0.19,0.01)$. Both give $s=0.6$, so the two mechanisms share the same QLDP parameter, $e^{\estar_\QLDP}=4$. With entanglement, however, the first has $e^{\estar_\EA}=5$ and $A_{\rm ent}=1.25$, while the second has $e^{\estar_\EA}=50$ and $A_{\rm ent}=12.5$; shrinking $w_Z$ further drives $A_{\rm ent}$ to infinity at the same QLDP parameter. The protection of a Pauli mechanism against entanglement attacks is thus governed by its rarest error, which its QLDP parameter cannot reveal. Anisotropic Pauli mechanisms are therefore vulnerable to entanglement attacks, and we recommend using more isotropic channels, such as the depolarizing channel.

For quantitative analysis of a more comprehensive list of mechanisms, including the damping-type mechanisms, we refer interested readers to Table~\ref{tab:channel-values} in Appendix~\ref{app:channels}.

\section{Experiments}\label{sec:exp}

We now test the entanglement attack and its defense in simulation and on a real quantum computer, answering three questions: Is the attack effective on today's noisy hardware (Sec.~\ref{sec:exp-motivating})? Does it scale to a computation that processes many sensitive bits at once (Sec.~\ref{sec:exp-attack})? Does EA-QDP restore the privacy guarantee (Sec.~\ref{sec:exp-defense})?

\emph{Setup.} Physical experiments run on the \texttt{ibm\_fez} QPU of the IBM quantum platform without error mitigation; simulations use the noiseless statevector simulator of Qiskit~\cite{qiskit} on a laptop (Intel i7-8750H CPU, 16\,GB memory). Every configuration is run for $2{,}048$ shots, i.e., repetitions of the circuit. Throughout this section, QDP refers to QDP within $\eta=1$, i.e., QLDP, the strongest guarantee without entanglement. Our code, circuits, and raw data are available at \url{https://anonymous.4open.science/r/EA-QDP_artifacts}.

\emph{Metric: from DP to accuracy.} To make privacy leakage observable, we let the adversary guess the sensitive bit and measure how often it is right. By Bayes' rule, $\vareps$-DP means that no outcome $o$ can change the adversary's odds on the bit by more than a factor of $e^{\vareps}$. We sample both values of the secret bits with equal probability, so that a blind guess is correct with probability $50\%$, and any guess $\hat x$ made upon outcome $o$ of an $\veps$-DP mechanism is correct with probability at most
\begin{equation}
  \Pr[x=\hat x|o]\le\frac{e^{\vareps}}{1+e^{\vareps}}.
  \label{eq:acc-bound}
\end{equation}
An accuracy above this bound is direct evidence that the mechanism violates $\vareps$-DP. Conversely, an observed accuracy $\mathrm{Acc.}$ yields the empirical estimate $\hat\vareps=\ln[\mathrm{Acc.}/(1-\mathrm{Acc.})]$, denoted by $\hat\vareps^*_\EA$ and $\hat\vareps^*_{\rm QDP}$ for adversaries with and without the hidden system, respectively. This methodology follows the empirical auditing of DP implementations~\cite{jagielski2020auditing,nasr2021adversary}. In all figures, solid markers and lines denote hardware results, and hollow markers and dotted lines denote simulated ones; error bars are $95\%$ confidence intervals for the sampling error of finitely many shots, but not for hardware noise.

\emph{Adversaries.} In every experiment, we compare two adversaries that face the same computation and the same mechanism, and differ only in the hidden system: one measures the released state alone, as assumed by QDP, and the other runs the entanglement attack. As is standard in DP, both know the mechanism but not the sensitive data. Any gap between their accuracies thus isolates the effect of entanglement.

\subsection{A Minimal Attack on the Motivating Example}\label{sec:exp-motivating}

\begin{figure}[!t]
  \centering
  \begin{tikzpicture}
    \begin{yquant}
      qubit {$R\colon|0\>$} r;
      qubit {$S\colon|0\>$} s;
      hspace {1mm} r, s;
      [name=prep] h r;
      [name=cx] cnot s | r;
      hspace {2mm} r, s;
      [name=ud] box {$U_D=X^x$} s;
      [name=mech] box {$\E_{XZ}$} s;
      box {$\mathcal M_{\rm Bell}$} (r, s);
      [value=cbit] settype r, s;
      hspace {1.5mm} r, s;
    \end{yquant}
    \node[draw=red!60!black,dashed,rounded corners,inner sep=1.2mm,fit=(prep)(cx),
      label={[font=\scriptsize,red!60!black]above:Entangled state preparation}] {};
    \node[draw=blue!65!black,dashed,rounded corners,inner sep=1.2mm,fit=(ud)(mech),
      label={[font=\scriptsize,blue!65!black]below:Quantum Computer}] {};
  \end{tikzpicture}\\[.1mm]
  \includegraphics[width=0.82\columnwidth]{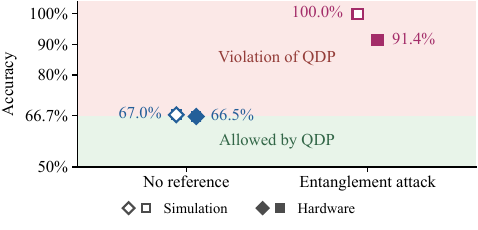}
  \caption{Realization of the motivating example. Top: the quantum circuit implementing the motivating example. Bottom: accuracy of the entanglement attack, pooled over both values of the sensitive bit $x$; the red region marks the accuracies that the $\ln 2$-QDP guarantee of $\E_{XZ}$ forbids.}
  \label{fig:exp-motivating}
\end{figure}
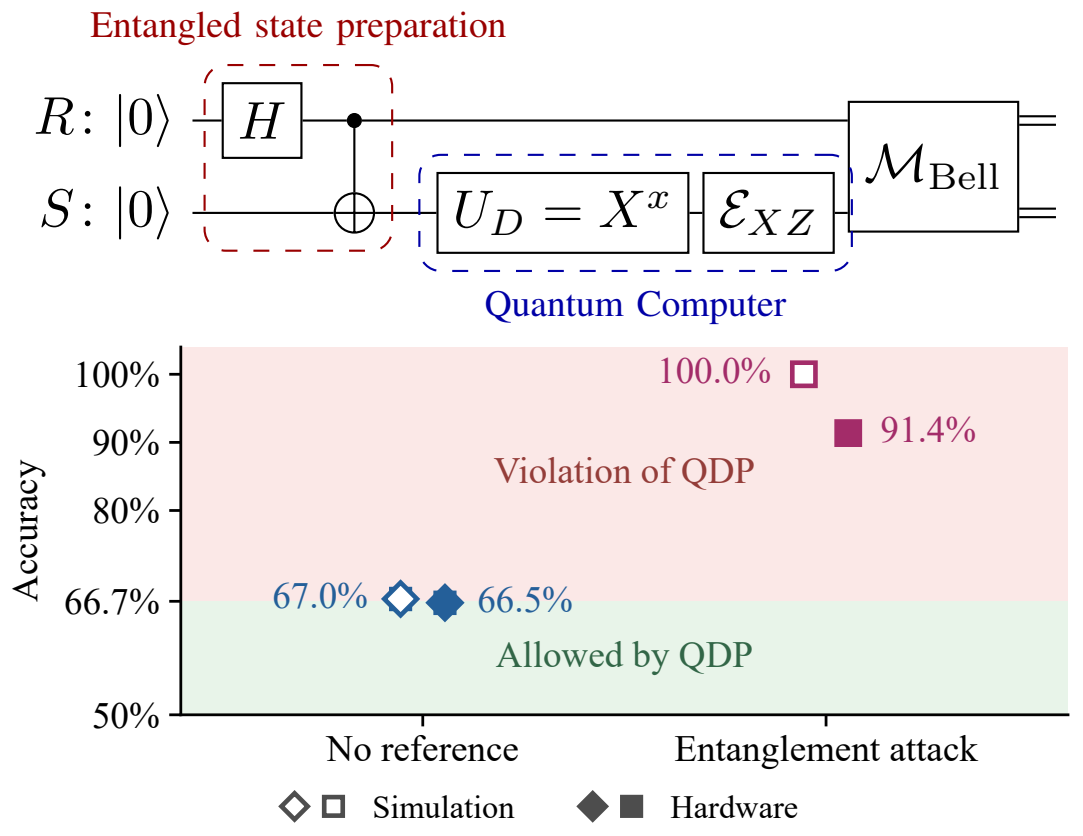

We first run the motivating example of Sec.~\ref{sec:motivating-example}, the smallest possible entanglement attack (Fig.~\ref{fig:exp-motivating}, top). The adversary prepares the Bell state $|\Phi^+\>$ on $R$ and $S$ with a Hadamard and a CNOT gate, keeps $R$, and submits $S$; the quantum computer encodes the sensitive bit $x$ by $U_D=X^x$ and applies the mechanism $\E_{XZ}$, which satisfies $\ln2$-QDP and thus permits an accuracy of at most $66.7\%$; finally, the adversary measures $R$ and the returned $S$ jointly in the Bell basis. By~\eqref{eq:motivating-zero-event}, the outcome $\Phi^-$ never occurs when $x=1$, and $\Psi^-$ never occurs when $x=0$; upon these outcomes, an ideal adversary recovers $x$ with certainty. These outcomes are exactly the zero entries of Table~\ref{tab:motivating}, which only the hidden qubit makes observable.

\emph{Results.} Fig.~\ref{fig:exp-motivating} (bottom) confirms this prediction. Without the hidden system, the adversary's accuracy stays at $67.0\%$ ($\hat\vareps^*_{\rm QDP}=0.71$), consistent with the QDP bound. With the hidden system, the entanglement attack reaches $100\%$ in simulation and $91.4\%$ on hardware, where the gap is due to device noise; the hardware accuracy amounts to $\hat\vareps^*_\EA=2.4$, more than three times the claimed $\estar_{\rm QDP}(\E_{XZ})=\ln2\approx0.69$. A mechanism that promises to leak no more than randomized response thus lets the adversary learn the bit almost for certain, using a single Bell pair.

\subsection{Scaling the Attack to Many Bits}\label{sec:exp-attack}

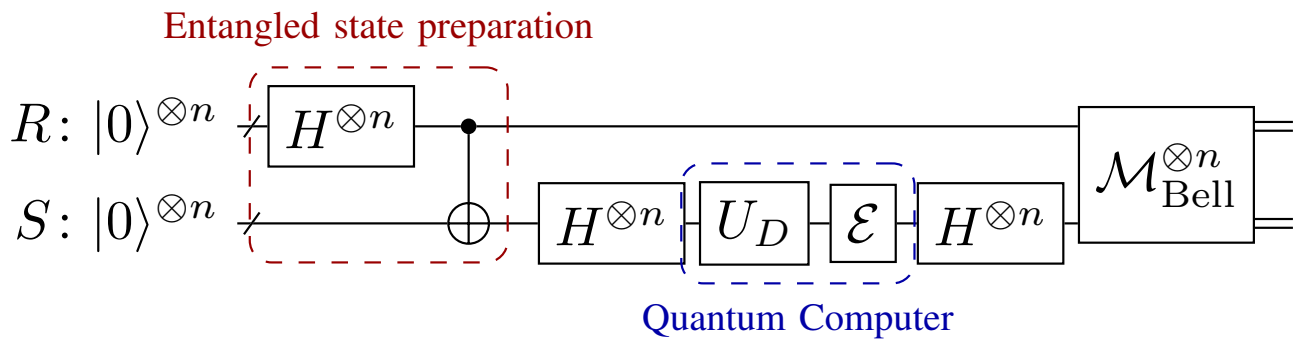
\begin{figure}[!t]
  \centering
  \resizebox{\columnwidth}{!}{%
  \begin{tikzpicture}
    \begin{yquant}
      qubit {$R\colon|0\>^{\otimes n}$} r;
      qubit {$S\colon|0\>^{\otimes n}$} s;
      slash r;
      slash s;
      hspace {1mm} r, s;
      [name=prep] box {$H^{\otimes n}$} r;
      [name=cx] cnot s | r;
      hspace {1mm} r, s;
      box {$H^{\otimes n}$} s;
      [name=ud] box {$U_D$} s;
      [name=mech] box {$\E$} s;
      box {$H^{\otimes n}$} s;
      box {$\mathcal M_{\rm Bell}^{\otimes n}$} (r, s);
      [value=cbit] settype r, s;
      hspace {1.5mm} r, s;
    \end{yquant}
    \node[draw=red!60!black,dashed,rounded corners,inner sep=1.2mm,fit=(prep)(cx),
      label={[font=\scriptsize,red!60!black]above:Entangled state preparation}] {};
    \node[draw=blue!65!black,dashed,rounded corners,inner sep=1.2mm,fit=(ud)(mech),
      label={[font=\scriptsize,blue!65!black]below:Quantum Computer}] {};
  \end{tikzpicture}%
  }
  \caption{The quantum circuit underlying the full-scale experiments (Sec.~\ref{sec:exp-attack} and Sec.~\ref{sec:exp-defense}). The second and third $H^{\otimes n}$ gate groups realize the Bernstein--Vazirani algorithm.}
  \label{fig:exp-circuit}
\end{figure}

We next attack a computation that processes $n$ sensitive bits $\v x=(x_1,\dots,x_n)$ at once, with $n$ up to $10$, given by the \emph{phase oracle}
\begin{equation}
  U_D|\v a\>=(-1)^{\v a\cdot\v x}|\v a\>,
\end{equation}
where $\v a\cdot\v x$ is taken modulo $2$. It mirrors a classical counting query, which counts the sensitive bits selected by the query $\v a$, and returns the count modulo $2$ as a phase. Unlike in the classical setting, a single such query suffices to learn all $n$ bits in the quantum setting: the Bernstein--Vazirani (BV) algorithm~\cite{bernstein1993quantum} applies $H^{\otimes n}$ before and after the oracle and the adversary will be able to read out the entire $\v x$ directly. Without the hidden system, the adversary runs BV on $S$ alone; in the entanglement attack, it additionally prepares a Bell pair $(R_i,S_i)$ for each bit $x_i$, and measures every pair in the Bell basis at the end (Fig.~\ref{fig:exp-circuit}).

\emph{Mechanisms and privacy budget.} We apply one copy of a mechanism to each submitted qubit. The first mechanism generalizes $\E_{XZ}$ with a tunable weight $q$ on the identity,
\begin{equation}
  \E_{XZ,q}(\rho)=q\rho+\tfrac{1-q}{2}(X\rho X+Z\rho Z),
\end{equation}
which recovers $\E_{XZ}$ at $q=1/3$. As a Pauli mechanism (Sec.~\ref{sec:amplification}), it has the QDP parameter $\ln[(1+s)/(1-s)]$ with $s=\max\{q,|2q-1|\}$, but an infinite EA-QDP parameter for every $q$, since the $Y$ branch is always missing. The second is the depolarizing mechanism $\D_p$, our recommended choice. We calibrate both $\E_{XZ,q}^{\otimes n}$ and $\D_p^{\otimes n}$ to a total QDP parameter of $3$, i.e., $3/n$ per qubit\footnote{It can be shown that $\estar_\EA(\E^{\otimes n})=n\ \estar_\EA(\E)$ for any $\E$, and $\estar_{\rm QDP}(\E^{\otimes n})=n\ \estar_{\rm QDP} (\E)$ holds for $\E=\E_{XZ,q}$ or $\mathcal D_p$ and $\eta=1$.}, and take the larger of the two solutions for $q$ for better utility\footnote{$\estar_{\rm QDP}(\E_{XZ,q})$ reaches its minimum $\ln 2$ at $q=1/3$. For large $n$, where the calibration $\estar_{\rm QDP}(\E_{XZ,q}^{\otimes n})=3$ is not achievable, we set $q=1/3$ and use the smallest achievable parameter instead.}. As in Sec.~\ref{sec:exp-motivating}, we report the accuracy with which the adversary recovers each bit $x_i$, conditioned on the informative measurement outcomes, together with the bound~\eqref{eq:acc-bound} given by the per-bit QDP parameter; since the total budget is fixed, the bound shrinks as $n$ grows.

\begin{figure}[!t]
  \centering
  \includegraphics[width=\columnwidth]{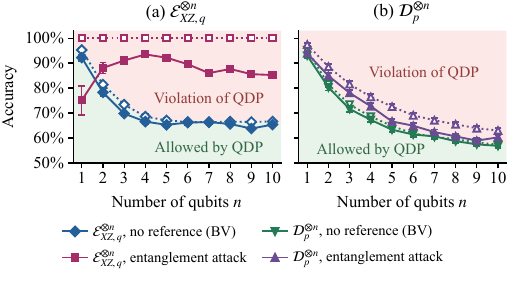}
  \caption{Accuracy of recovering the sensitive bits against $\E_{XZ,q}^{\otimes n}$ (a) and $\D_p^{\otimes n}$ (b), both calibrated to a total QDP parameter of $3$. The red region marks the accuracies that the QDP parameter of each bit forbids. Solid lines with filled markers show hardware results, and dotted lines with hollow markers show simulated ones.}
  \label{fig:exp-attack}
\end{figure}

\emph{Results.} Fig.~\ref{fig:exp-attack} shows a clear separation between the two adversaries. Without the hidden system, the BV adversary attains exactly what QDP permits: its simulated accuracy lies on the boundary of the allowed region at every $n$, and its hardware accuracy are slightly lower. With the hidden system, the same mechanisms leak beyond the QDP bound, by the amount that EA-QDP predicts. As the two adversaries differ only in the hidden system, this excess leakage is caused by entanglement alone. For $\E_{XZ,q}^{\otimes n}$, whose EA-QDP parameter is infinite, the entanglement attack recovers every guessed bit correctly in simulation, i.e., $\hat\vareps^*_\EA=+\infty$, and on hardware it reaches $93.5\%$ at $n=4$ and $85.2\%$ at $n=10$, against the $66.7\%$ that QDP permits. These amount to $\hat\vareps^*_\EA=2.67$ and $1.75$, far above both the claimed $\estar_{\rm QDP}=0.75$ and $0.69$ and the empirical $\hat\vareps^*_{\rm QDP}=0.69$ and $0.64$ without the hidden system. Device noise accounts for the gap to $100\%$, and also pulls the hardware accuracy below the bound at $n=1$, where informative outcomes are rare. Despite the stronger noise of larger circuits, the attack stays above the QDP bound on hardware for every $n\ge2$, and far above it for $n\ge3$.

For $\D_p^{\otimes n}$, whose entanglement amplification is small (Sec.~\ref{sec:amplification}), the gap is much milder, yet the attack still exceeds the QDP bound on hardware for every $n\ge2$, e.g., $60.3\%$ against $57.4\%$ at $n=10$, i.e., $\hat\vareps^*_\EA=0.42$, above both the claimed $\estar_{\rm QDP}=0.30$ and the empirical $\hat\vareps^*_{\rm QDP}=0.28$. In simulation, the attack attains exactly the EA-QDP parameter of the mechanism, e.g., $\hat\vareps^*_\EA=\estar_\EA=0.53$ at $n=10$. Calibrated by QDP, neither mechanism keeps its promise against the entanglement attack, and the size of the violation follows the EA-QDP parameter: unbounded for $\E_{XZ,q}$, and mild for $\D_p$.

\subsection{Defense Calibrated by EA-QDP}\label{sec:exp-defense}

Our theory suggests a two-step defense: choose a mechanism with small entanglement amplification, and calibrate it by its EA-QDP parameter rather than its QDP parameter. We test this defense with the same computation, circuit, and attack as in Sec.~\ref{sec:exp-attack}.

\emph{Mechanisms and privacy budget.} We spend a total EA-QDP budget of $3$ in two ways: $\D_p^{\otimes n}$ as before, now with an EA-QDP parameter of $3/n$ per qubit, and the global depolarizing mechanism $\D_p^{\rm glob}(\rho)=p\rho+(1-p)I/2^n$, which replaces the whole register with noise at once, and whose EA-QDP parameter $\ln[1+4^np/(1-p)]$ (Sec.~\ref{sec:amplification}) we set to $3$. For each bit, we bound the accuracy by~\eqref{eq:acc-bound} with the EA-QDP parameter of the mechanism that the bit experiences.

\begin{figure}[!t]
  \centering
  \includegraphics[width=\columnwidth]{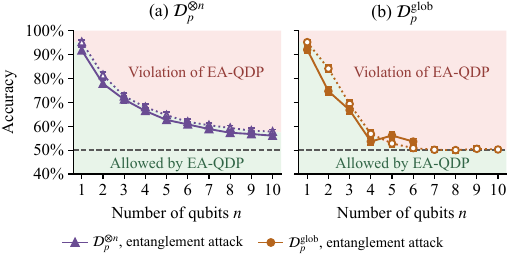}
  \caption{Accuracy of the entanglement attack against $\D_p^{\otimes n}$ (a) and $\D_p^{\rm glob}$ (b), both calibrated to a total EA-QDP parameter of $3$. The red region marks the accuracies that the EA-QDP parameter of the mechanism acting on each bit forbids.}
  \label{fig:exp-defense}
\end{figure}

\emph{Results.} Fig.~\ref{fig:exp-defense} shows that the defense restores the privacy guarantee. For $\D_p^{\otimes n}$, every hardware result stays within the region that EA-QDP allows, and the simulated accuracy lies exactly on its boundary: the attack attains what EA-QDP permits, and no more. Compared with Sec.~\ref{sec:exp-attack}, both calibrations claim the same per-bit parameter of $0.30$ at $n=10$, yet on hardware the attack achieves $\hat\vareps^*_\EA=0.42$ against the calibration by QDP, breaking the claim, and only $0.24$ against the calibration by EA-QDP. The price is stronger noise: each qubit keeps its input with probability $p\approx0.080$ instead of $0.149$.

\emph{Cost of the defense.} The stronger noise is the minimum price of protection against entanglement attacks. Each qubit of $\D_p^{\otimes n}$ preserves a pure input with fidelity $(1+p)/2$, which drops only from $0.574$ under the calibration by QDP to $0.540$ under the calibration by EA-QDP at $n=10$. By Theorem~\ref{thm:optimal-utility}, $0.540$ is the highest fidelity that any qubit mechanism can achieve with the same EA-QDP parameter, so no other mechanism offers a better tradeoff. In contrast, no calibration can rescue $\E_{XZ,q}$: its EA-QDP parameter is infinite for every $q$, so even the noisiest $\E_{XZ,q}$ remains vulnerable. A sound defense thus requires both steps, choosing the right mechanism and calibrating it by EA-QDP.

\begin{table}[!t]
  \centering
  \caption{Claimed versus empirical per-bit privacy parameters (multi-qubit results at $n=10$).}
  \label{tab:exp-summary}
  \footnotesize
  \setlength{\tabcolsep}{3.2pt}
  \newcommand{\leak}[1]{\textcolor{red!60!black}{$\bm{#1}$}}
  \begin{tabular}{@{}cllcccc@{}}
    \toprule
    & & & & & \multicolumn{2}{c}{$\hat\vareps^*_\EA$} \\
    \cmidrule(l){6-7}
    Sec. & Mechanism & Calibrated by & Claimed & $\hat\vareps^*_{\rm QDP}$ & HW & Sim. \\
    \midrule
    \ref{sec:exp-motivating} & $\E_{XZ}$ & QDP & $0.69$ & $0.71$ & \leak{2.4} & \leak{+\infty} \\
    \ref{sec:exp-attack} & $\E_{XZ,q}^{\otimes 10}$ & QDP & $0.69$ & $0.64$ & \leak{1.75} & \leak{+\infty} \\
    \ref{sec:exp-attack} & $\D_p^{\otimes 10}$ & QDP & $0.30$ & $0.28$ & \leak{0.42} & \leak{0.53} \\
    \ref{sec:exp-defense} & $\D_p^{\otimes 10}$ & EA-QDP & $0.30$ & -- & {\textcolor{green!50!black}{$\bm{0.24}$}} & {\textcolor{green!50!black}{$\bm{0.30}$}} \\
    \bottomrule
  \end{tabular}
\end{table}

For $\D_p^{\rm glob}$, the simulated accuracy again follows the bound, which quickly approaches the $50\%$ of a blind guess, since the same budget now covers all $n$ bits at once. On hardware, the accuracy stays within the bound for $n\le4$, but slightly exceeds it for larger $n$, e.g., $56.1\%$ against $51.8\%$ at $n=5$. This excess stems from the hardware implementation rather than the theory: we realize $\D_p^{\rm glob}$ with probabilistic controlled-SWAP gates on all $n$ qubits, whose imperfections let part of the input survive, and a bound this close to $50\%$ leaves no room for such residual leakage. Moreover, $\D_p^{\rm glob}$ needs far stronger noise to reach the same EA-QDP parameter: at $n=5$, each qubit retains its input with probability $p\approx0.018$, against $0.17$ under $\D_p^{\otimes n}$, so the output is of little use for further computation. Both reasons make $\D_p^{\otimes n}$ the more practical choice.

\emph{Lessons from hardware.} First, in most configurations, device noise lowers the accuracy of the attack relative to simulation, so the violations of QDP observed on hardware understate the power of the attack, which will approach its simulated accuracy as quantum hardware improves. Second, an imperfect implementation of the mechanism itself can leak beyond its theoretical guarantee, as $\D_p^{\rm glob}$ shows. Privacy mechanisms on quantum hardware should therefore be implemented faithfully, and simple qubit-wise mechanisms such as $\D_p^{\otimes n}$ are more robust in this respect.

\emph{Summary.} Table~\ref{tab:exp-summary} summarizes our results. Across all experiments, the measured leakage agrees with our theory. Calibrated by QDP, mechanisms leak beyond their claimed guarantees, by as much as their EA-QDP parameters predict; calibrated by EA-QDP, the depolarizing mechanism withstands the entanglement attack, with a bound that is attained in simulation and respected on hardware.

\section{Related Work}\label{sec:related}

\emph{Quantum differential privacy.} Since its introduction~\cite{zhou2017Differential}, quantum differential privacy has been studied from the information-theoretic perspective~\cite{hirche2023Quantum}, through gentle measurements~\cite{aaronson2019Gentle} and unified algorithmic frameworks~\cite{angrisani2023Unifying}, and in the local model~\cite{angrisani2025quantum,guan2025Optimal}, with applications to quantum classifiers~\cite{du2021Quantum}, recommendation systems~\cite{li2026Differential}, and sensor networks~\cite{spencer2026Differentially}. Tools have also been developed to detect privacy violations~\cite{guan2023Detectinga} and to optimize the privacy--utility tradeoff, where the depolarizing mechanism is optimal~\cite{guan2025Optimal,nuradha2026PrivacyUtility}. All these frameworks, however, assume that the adversary measures only the released state, and evaluating their privacy parameters generally requires non-convex optimization~\cite{guan2023Detectinga,guan2025Optimal}. EA-QDP removes this assumption while keeping existing notions as special cases: without a hidden system, it reduces to QLDP and hence implies QDP. Unlike them, its parameter is computed exactly by a convex SDP. Moreover, the optimality of the depolarizing mechanism carries over to EA-QDP, and a finiteness criterion previously stated for QLDP~\cite{guan2025Optimal} turns out to characterize EA-QDP instead (Remark~\ref{remark:erratum}). EA-QDP thus extends existing guarantees to entangled adversaries without sacrificing computability or utility.

\emph{Entanglement in quantum privacy.} Closest to our work, several recent studies examine how entanglement interacts with quantum privacy. Entanglement can make individually private channels completely lose privacy when they are implemented jointly~\cite{alabi2026When}; QLDP mechanisms with sufficiently small privacy parameters must break entanglement~\cite{bhalerao2026High}; and entanglement, viewed as a constraint on the input states, can even improve privacy~\cite{wang2026How}. Quantum pufferfish privacy~\cite{nuradha2024Quantum} further sketches a variant with reference systems, but allows the two inputs to differ on the reference system, although an adversary's hidden system never touches the sensitive data. None of these works models entanglement as a capability of an adversary, so none of them captures entanglement attacks. Breaking entanglement alone does not stop such attacks either: the mechanism $\E_{XZ}$ in Sec.~\ref{sec:motivating-example} is entanglement-breaking, yet completely broken. EA-QDP models this adversary precisely, by requiring the two inputs to agree on the hidden system, and quantifies the resulting leakage through the entanglement amplification. Our work thus exposes the adversarial side of entanglement, complementing the view in~\cite{wang2026How} that entanglement can improve privacy.

\emph{Attacks on quantum systems and privacy auditing.} Security proofs in quantum cryptography have long accounted for adversaries that entangle their probes with the transmitted signals and measure them jointly, such as coherent attacks on quantum key distribution~\cite{coherent_attack,quantum_crypto_advances}. Quantum computers additionally face implementation-level attacks, such as high-energy state attacks~\cite{xu2023securing} and power side channels~\cite{xu2023exploration}. For DP, auditing instantiates concrete adversaries to test claimed guarantees empirically~\cite{jagielski2020auditing,nasr2021adversary}, and violations of QDP can be detected numerically~\cite{guan2023Detectinga}. However, these attacks target cryptographic keys or hardware, and existing quantum privacy audits consider only adversaries without entanglement. Our entanglement attack brings the entangled adversary of quantum cryptography to quantum DP, and requires nothing beyond the standard user interface. Our audit on IBM quantum hardware shows that it breaks the QDP bound by a wide margin, whereas mechanisms calibrated by EA-QDP withstand it. EA-QDP is thereby validated not only in theory but also on physical hardware.

\emph{Entanglement-assisted discrimination and communication.} Technically, our analysis connects to quantum information theory, where entanglement is known to help distinguish quantum channels: every entangled state is useful in some channel discrimination task~\cite{piani2009All}, and a reference system of the input dimension suffices for optimal discrimination~\cite{puzzuoli2017Ancilla}.
Relatedly, the capacity of entanglement-assisted postselected communication is characterized by the projective mutual information $I_\Omega(\E)$ of a channel, which is also computed by an SDP~\cite{ji2024Postselected}. These results optimize discrimination or communication performance, whereas DP requires a worst-case bound on the likelihood ratio between any two inputs that agree on the reference system. We establish the privacy counterparts: a reference system of the input dimension suffices for the adversary (Theorem~\ref{thm:dim-bound}), and $\estar_\EA$ is computed by an SDP of the same form as in~\cite{ji2024Postselected}, with $\estar_\EA(\E)=(\ln 2)\cdot I_\Omega(\E)$. This identity links privacy leakage under entanglement attacks to entanglement-assisted communication, suggesting that tools developed for one may be transferred to the other.

\section{Conclusion}\label{sec:conclusion}
In the work we proposed EA-QDP, a privacy framework that remains valid under the entanglement attacks. In particular, we showed that an adversary holding a hidden entangled reference can completely break mechanisms that satisfy quantum differential privacy, proposed EA-QDP to capture and counteract this threat, showed that its optimal parameter is computable by an SDP, and found that well-chosen mechanisms such as the depolarizing channel can be resilient to such attacks, as confirmed on quantum hardware.

\bibliographystyle{IEEEtran}
\bibliography{libpaper}

\appendices

\section*{AI Tool Usage Statement}

The authors used GPT-5.6, GPT-6 and GPT-6.1 in the development of this work, and used GPT-6.1 and Claude Opus 5.5 for editorial purposes in this manuscript. All mathematical proofs, analysis, citations and the final text were reviewed and validated by the authors, who take full responsibility for the final work.
\section{Deferred Proofs}\label{app:proofs}

This appendix collects the proofs deferred from the main text.
\subsection{Proof of Theorem~\ref{thm:dim-bound}}\label{app:proof-dim-bound}
The ``only if'' direction is immediate. For the ``if'' direction, we move every violation on a large reference system $R$ to a reference system $r$ with $\dim(\H_r)\le d$. A violation is always detected by testing a single vector $|W\>$ on $R\ts S$, and such a vector involves at most $d$ directions of $R$, so we can cut $R$ down to these directions.

\emph{A rank-one witness.} Suppose that states $\rho_{RS},\sig_{RS}$ with $\rho_R=\sig_R$, a measurement $\M$ and an event $E$ violate~\eqref{eq:ea-qdp}, i.e., $\Tr(M_E\delta)>0$ for $M_E:=\sum_{k\in E}M_k\ge0$ and
\begin{equation}
  \delta:=(\I_R\ts\E)(\rho_{RS}-e^{\vareps}\sig_{RS}).
\end{equation}
Then the Hermitian operator $\delta$ has a positive eigenvalue, and a unit eigenvector $|W\>$ for it satisfies $\<W|\delta|W\>>0$.

\emph{Cutting $R$ down.} By the Schmidt decomposition (the singular value decomposition of the coefficient matrix of $|W\>$), $|W\>=\sum_{i=1}^m\sqrt{\lambda_i}\,|a_i\>|b_i\>$ with orthonormal $\{|a_i\>\}\subseteq\H_R$ and $\{|b_i\>\}\subseteq\H_S$, so $m\le d$. Let $r$ be an $m$-dimensional system with basis $\{|i\>\}_{i=1}^m$, let $V:=\sum_i|a_i\>\<i|$ embed $\H_r$ into $\H_R$, so that $V^\dagger V=I_r$, and let $\mathcal C(X):=(V^\dagger\ts I)X(V\ts I)$ cut an operator $X$ on $R\ts S$ down to $r\ts S$. Then $|w\>:=(V^\dagger\ts I)|W\>$ is a unit vector with $(V\ts I)|w\>=|W\>$. As $\mathcal C$ acts on the reference system only, it commutes with the mechanism, i.e., $(\I_r\ts\E)\circ\mathcal C=\mathcal C\circ(\I_R\ts\E)$, and hence
\begin{equation}
  \<w|(\I_r\ts\E)(\mathcal C(X))|w\>=\<W|(\I_R\ts\E)(X)|W\>.
  \label{eq:app-compress}
\end{equation}

\emph{The violation on $r$.} Let $\rho_{rS}:=\mathcal C(\rho_{RS})/N$ and $\sig_{rS}:=\mathcal C(\sig_{RS})/N$ with the \emph{same} factor $N:=\Tr(VV^\dagger\rho_R)=\Tr(VV^\dagger\sig_R)$; this is where $\rho_R=\sig_R$ enters. We have $N>0$, since otherwise the positive operators $\mathcal C(\rho_{RS})$ and $\mathcal C(\sig_{RS})$ have zero trace and vanish, and~\eqref{eq:app-compress} with $X=\rho_{RS}-e^{\vareps}\sig_{RS}$ would give $\<W|\delta|W\>=0$. Hence $\rho_{rS},\sig_{rS}$ are states, and they are neighbors, as $\rho_r=V^\dagger\rho_RV/N=\sig_r$. By~\eqref{eq:app-compress}, $\<w|(\I_r\ts\E)(\rho_{rS}-e^{\vareps}\sig_{rS})|w\>=\<W|\delta|W\>/N>0$, so the measurement $\{|w\>\<w|,I-|w\>\<w|\}$ violates~\eqref{eq:ea-qdp} on $r$, where $\dim(\H_r)=m\le d$.
\hfill$\square$

\subsection{Proof of Theorem~\ref{thm:cp-sandwich}}\label{app:proof-cp-sandwich}

\emph{Sufficiency.} The chain of inequalities after Theorem~\ref{thm:cp-sandwich} holds for every reference system $R$ by Choi's theorem~\cite{choi1975completely}: a linear map $\Phi$ is \emph{completely positive} (CP), i.e., $(\I_R\ts\Phi)(Z)\ge0$ for every $R$ and every $Z\ge0$, if and only if $J_\Phi\ge0$. Writing $\Phi_1\le_{\rm CP}\Phi_2$ if $\Phi_2-\Phi_1$ is CP, and using that $J_\Phi$ is linear in $\Phi$ with $J_{\mathcal R_\omega}=I_R\ts\omega$, we see that~\eqref{eq:cp-sandwich} is equivalent to
\begin{equation}
  \E\le_{\rm CP}\mathcal R_\omega\le_{\rm CP}e^{\vareps}\E.
  \label{eq:app-cp-order}
\end{equation}
Taking the trace of the resulting operator inequalities against $M_E:=\sum_{k\in E}M_k\ge0$ gives~\eqref{eq:ea-qdp}. Since both constraints in~\eqref{eq:cp-sandwich} are linear in $(e^{\vareps},\omega)$, the SDP~\eqref{eq:sdp} follows from the equivalence.

\emph{Necessity.} Suppose that no $\omega$ satisfies~\eqref{eq:cp-sandwich}; we construct an attack with $\dim(\H_R)=d$ that violates~\eqref{eq:ea-qdp}. The idea is to separate, by a hyperplane, the gaps $I_R\ts\omega-J_\E$ and $e^{\vareps}J_\E-I_R\ts\omega$ of all candidate $\omega$ from the pairs of positive semidefinite matrices. This hyperplane is itself the attack: its normal vector is a pair of positive operators with equal marginals, which become two admissible inputs after normalization and a swap of $R$ and $S$, while the measurement is fixed in advance and independent of $\E$, namely the projection onto the maximally entangled state $|\Omega\>$.

\emph{Separation.} Let $\mathcal V$ be the real vector space of pairs $(U,V)$ of Hermitian operators on $\H_R\ts\H_S$; every linear functional on $\mathcal V$ has the form $\ell(U,V)=\Tr(PU)+\Tr(QV)$ for Hermitian $P,Q$. We collect the gaps of all candidates in $\mathcal A$ and the positive semidefinite pairs in $\mathcal B$:
\begin{gather}
  \mathcal A:=\big\{\big(I_R\ts\omega-J_\E,\ e^{\vareps}J_\E-I_R\ts\omega\big):\ \omega=\omega^\dagger\big\},\nonumber\\
  \mathcal B:=\{(U,V):\ U\ge0,\ V\ge0\}.\nonumber
\end{gather}
A point of $\mathcal A\cap\mathcal B$ corresponds to a Hermitian $\omega$ satisfying~\eqref{eq:cp-sandwich}, which is automatically positive since $I_R\ts\omega\ge J_\E\ge0$; hence $\mathcal A\cap\mathcal B=\varempty$. Both sets are convex, and $\mathcal B-\mathcal A:=\{b-a\mid a\in\mathcal A,\ b\in\mathcal B\}$ is closed.\footnote{Elements of $\mathcal B-\mathcal A$ have the form $(U-I_R\ts\omega+J_\E,\ V+I_R\ts\omega-e^{\vareps}J_\E)$ with $U,V\ge0$. Take a convergent sequence of such elements. If $(\omega_n)$ is unbounded, pass to a subsequence with $\|\omega_n\|\to\infty$ and $\omega_n/\|\omega_n\|\to\omega^*$, $\|\omega^*\|=1$; dividing by $\|\omega_n\|$ shows $I_R\ts\omega^*=\lim U_n/\|\omega_n\|\ge0$ and $-I_R\ts\omega^*=\lim V_n/\|\omega_n\|\ge0$, a contradiction. Hence $(\omega_n)$, and with it $(U_n)$ and $(V_n)$, is bounded, and a convergent subsequence shows that the limit lies in $\mathcal B-\mathcal A$.} Closedness is what makes the final inequality strict: by the separating hyperplane theorem~\cite[Sec.~2.5.1]{boyd2004convex}, a closed convex set that does not contain $0$ lies strictly on one side of a hyperplane through $0$. Applied to $\mathcal B-\mathcal A$, it gives Hermitian $P,Q$ with
\begin{equation}
  \ell(a)<\ell(b)\qquad\forall\,a\in\mathcal A,\ b\in\mathcal B.
  \label{eq:app-strict-sep}
\end{equation}

\emph{Reading off the inputs.} Since $\mathcal B$ is closed under scaling by $\lambda\ge0$, \eqref{eq:app-strict-sep} forces $\ell\ge0$ on $\mathcal B$, as otherwise $\ell(\lambda b)$ would fall below $\ell(a)$ for large $\lambda$; testing $b=(|\psi\>\<\psi|,0)$ and $b=(0,|\psi\>\<\psi|)$ gives $P,Q\ge0$. Taking $b=0$ shows $\ell<0$ on $\mathcal A$, and since $\Tr[(I_R\ts\omega)P]=\Tr(\omega\Tr_R P)$, this reads
\begin{equation}
  \Tr\big[\omega(\Tr_R P-\Tr_R Q)\big]-\Tr(J_\E P)+e^{\vareps}\Tr(J_\E Q)<0
  \nonumber
\end{equation}
for every Hermitian $\omega$. The first term is unbounded, by taking $\omega$ a large multiple of $\Tr_R P-\Tr_R Q$, unless $\Tr_R P=\Tr_R Q$. Hence $P$ and $Q$ have equal marginals, exactly as two admissible inputs, and
\begin{equation}
  \Tr(J_\E P)>e^{\vareps}\Tr(J_\E Q).
  \label{eq:app-certificate}
\end{equation}
In particular, $P\ne0$, so $t:=\Tr P=\Tr Q>0$, where the traces agree because the marginals do.

\emph{The attack.} The marginals of $P$ and $Q$ agree on $S$ rather than on $R$, so we swap the two systems. Let $F$ be the swap operator, $F|i\>_R|j\>_S=|j\>_R|i\>_S$, let $\top$ denote the transpose in the computational basis, and let the inputs be $\rho_{RS}:=F P^\top F/t$ and $\sig_{RS}:=F Q^\top F/t$. Both are states, since the transpose and the conjugation by the unitary $F$ preserve positivity and trace, and $\rho_R=(\Tr_R P)^\top/t=\sig_R$. The adversary measures $\{M,I-M\}$ with $M:=|\Omega\>\<\Omega|/d$ (so $0\le M\le I$, as $\<\Omega|\Omega\>=d$), which tests whether $R\ts S$ is in the maximally entangled state; for a qubit, $M$ is the outcome $\Phi^+$ of the Bell measurement in Sec.~\ref{sec:motivating-example}. A direct computation shows that $\<\Omega|(\I_R\ts\E)(FX^\top F)|\Omega\>=\Tr(J_\E X)$ for every operator $X$.\footnote{Expand $J_\E=\sum_{i,j}|i\>\<j|\ts\E(|i\>\<j|)$ and use $\<a,b|FX^\top F|c,f\>=\<f,c|X|b,a\>$: both sides equal $\sum_{i,j,k,l}\<k|\E(|i\>\<j|)|l\>\<j,l|X|i,k\>$.} Hence $\Tr[M(\I_R\ts\E)(\rho_{RS})]=\Tr(J_\E P)/(d\,t)$ and $\Tr[M(\I_R\ts\E)(\sig_{RS})]=\Tr(J_\E Q)/(d\,t)$, so~\eqref{eq:app-certificate} violates~\eqref{eq:ea-qdp} for the outcome $M$.
\hfill$\square$

\subsection{Proof of Theorem~\ref{thm:ea-finite}}\label{app:proof-ea-finite}

The proof compares two inputs with the same marginal $I_R/d$ on $R$: the entangled input $|\Omega\>\<\Omega|/d$, with output $J_\E/d$, and the maximally mixed input $I_R\ts I_S/d^2$, with output $I_R\ts\E(I_S)/d^2>0$. An eigenvector of $J_\E$ with eigenvalue zero gives an outcome that the former never produces but the latter does; a positive definite $J_\E$, in turn, lies between two multiples of $I_R\ts I_S$, the Choi matrix of a replacement map.

\emph{Necessity.} Suppose that $J_\E$ has a zero eigenvalue, and let $M\neq0$ be the projector onto the eigenvectors with eigenvalue zero, so that $\Tr(MJ_\E)=0$. Take $\dim(\H_R)=d$, the neighbors $\rho_{RS}:=I_R\ts I_S/d^2$ and $\sig_{RS}:=|\Omega\>\<\Omega|/d$, and the measurement $\{M,I-M\}$. The outcome $M$ occurs with probability $\Tr[M(I_R\ts\E(I_S))]/d^2>0$ on $\rho_{RS}$, since $I_R\ts\E(I_S)>0$ and $M\neq0$, but with probability $\Tr(MJ_\E)/d=0$ on $\sig_{RS}$. Hence no finite $\vareps$ satisfies~\eqref{eq:ea-qdp}, and $\vareps^*_{\rm EA}(\E)=+\infty$.

\emph{Sufficiency.} Suppose that $J_\E>0$ with smallest and largest eigenvalues $0<\lambda_{\min}\le\lambda_{\max}$. Then $\omega:=\lambda_{\max}I_S$ gives $I_R\ts\omega=\lambda_{\max}I$, and multiplying $\lambda_{\min}I\le J_\E$ by $\lambda_{\max}/\lambda_{\min}$ yields $J_\E\le I_R\ts\omega\le(\lambda_{\max}/\lambda_{\min})J_\E$. By Theorem~\ref{thm:cp-sandwich}, $\E$ is $\ln(\lambda_{\max}/\lambda_{\min})$-EA-QDP, so $\vareps^*_{\rm EA}(\E)<+\infty$.
\hfill$\square$

\subsection{Proof of Theorem~\ref{thm:ea-values}}\label{app:proof-ea-values}
A Pauli matrix applied to half of a Bell pair yields another Bell state, just as the computation for $x=1$ turned $\Phi^+$ into $\Psi^+$ in Sec.~\ref{sec:motivating-example}. Hence $J_\E$ is diagonal in the Bell basis with the weights $w_\alpha$ on its diagonal, while a replacement map cannot tell the Bell states apart; this forces $\g^*=w_{\max}/w_{\min}$, where $w_{\max}$ and $w_{\min}$ denote the largest and smallest weights. Concretely, with $|\Phi^+\>=|\Omega\>/\sqrt2$, let $|\Phi_\alpha\>:=(I_R\ts\sigma_\alpha)|\Phi^+\>$ for $\alpha\in\{I,X,Y,Z\}$. These are the Bell states $\Phi^+,\Psi^+,\Psi^-,\Phi^-$ (up to a phase for $\alpha=Y$) and form an orthonormal basis, so $I_R\ts I_S=\sum_\alpha|\Phi_\alpha\>\<\Phi_\alpha|$, and by linearity $J_\E=2\sum_\alpha w_\alpha|\Phi_\alpha\>\<\Phi_\alpha|$.

\emph{Upper bound.} If $w_{\min}>0$, then $\omega:=2w_{\max}I_S$ gives $J_\E\le I_R\ts\omega\le(w_{\max}/w_{\min})J_\E$, by comparing the diagonal entries $2w_\alpha\le2w_{\max}\le(w_{\max}/w_{\min})\,2w_\alpha$. Hence $\g^*\le w_{\max}/w_{\min}$ in~\eqref{eq:sdp}.

\emph{Lower bound.} Let $(\g,\omega)$ be feasible in~\eqref{eq:sdp}. Every Bell state has marginal $I/2$ on $S$, so $\<\Phi_\alpha|I_R\ts\omega|\Phi_\alpha\>=\Tr(\omega I/2)=\Tr(\omega)/2$ for all $\alpha$. Taking $\<\Phi_\alpha|\cdot|\Phi_\alpha\>$ of both inequalities in~\eqref{eq:sdp} at the Bell states of largest and smallest weight gives
\begin{equation}
  2w_{\max}\le\Tr(\omega)/2\le\g\cdot2w_{\min}.
\end{equation}
If $w_{\min}>0$, then $\g\ge w_{\max}/w_{\min}$, so $\g^*=w_{\max}/w_{\min}$, and Theorem~\ref{thm:cp-sandwich} gives the claimed value. If $w_{\min}=0$, the chain reads $2w_{\max}\le0$, which is impossible since $w_{\max}\ge1/4$; hence~\eqref{eq:sdp} is infeasible and $\vareps^*_{\rm EA}(\E)=+\infty$.

\emph{Special cases.} Each of $\E_X$, $\E_Y$, $\E_Z$, and $\E_{XZ}$ has a zero weight, so $\vareps^*_{\rm EA}=+\infty$. For $\D_p$, $w_I=(1+3p)/4\ge w_X=w_Y=w_Z=(1-p)/4$, which gives the ratio $(1+3p)/(1-p)$ for $p<1$ and three zero weights at $p=1$.
\hfill$\square$

\subsection{Proof of Theorem~\ref{thm:optimal-utility}}\label{app:proof-optimal-utility}

The proof symmetrizes $\E$: rotating it by a unitary preserves its privacy and utility, and averaging over all rotations preserves privacy and can only improve the worst-case utility. The average treats all directions alike, so it is a depolarizing channel, and among these, $\D_{p^*}$ is the most useful one within the privacy budget. Throughout, $\g:=e^{\vareps}$.

\emph{Privacy.} For a unitary $U$, let $\E_U(\rho):=U^\dagger\E(U\rho U^\dagger)U$ be the mechanism rotated by $U$, and let $\bar\E:=\int\E_U\,dU$ be its average over the uniform (Haar) measure on unitaries. By Theorem~\ref{thm:cp-sandwich}, some $\omega\ge0$ satisfies $\E\le_{\rm CP}\mathcal R_\omega\le_{\rm CP}\g\E$, as in~\eqref{eq:app-cp-order}. Rotation preserves the CP order, since composing with unitary conjugations preserves complete positivity, and it maps $\mathcal R_\omega$ to $\mathcal R_{U^\dagger\omega U}$; averaging preserves it as well, since CP maps form a convex cone. Hence $\bar\E\le_{\rm CP}\mathcal R_{\bar\omega}\le_{\rm CP}\g\bar\E$ with $\bar\omega:=\int U^\dagger\omega U\,dU$, and $\bar\E$ is $\vareps$-EA-QDP by Theorem~\ref{thm:cp-sandwich}.

\emph{Structure.} Replacing $U$ by $UV$ leaves the Haar average unchanged, so $\bar\E$ is \emph{covariant}: $\bar\E(V\rho V^\dagger)=V\bar\E(\rho)V^\dagger$ for every unitary $V$. For a pure state $\psi$, every $V$ with $V|\psi\>=|\psi\>$ thus commutes with $\bar\E(\psi)$. Such $V$ act as arbitrary unitaries on the orthogonal complement of $|\psi\>$, which forces $\bar\E(\psi)=a\psi+b(I-\psi)$, and covariance makes $a,b$ independent of $\psi$. By linearity and trace preservation, $\bar\E=\D_{p'}$ with $p'=a-b$, which may be negative.

\emph{Utility.} Write $\psi_U:=U\psi U^\dagger$. For every pure state $\psi$, linearity gives the first line below, and the convexity and unitary invariance of the trace distance give the second:
\begin{align}
  \<\psi|\D_{p'}(\psi)|\psi\>&=\int\Tr\big[\psi_U\,\E(\psi_U)\big]\,dU\ge F(\E),\nonumber\\
  T\big(\D_{p'}(\psi),\psi\big)&\le\int T\big(\E(\psi_U),\psi_U\big)\,dU\le1-\hat T(\E).\nonumber
\end{align}
Hence $F(\E)\le F(\D_{p'})$ and $\hat T(\E)\le\hat T(\D_{p'})$.

\emph{Optimizing over depolarizing channels.} A direct calculation gives $F(\D_p)=\hat T(\D_p)=((d-1)p+1)/d$, which increases with $p$. By~\eqref{eq:dep-values}, $e^{\estar_\EA(\D_p)}$ increases with $p\in[0,1)$, equals $\g$ at $p=p^*$, and is $+\infty$ at $p=1$.\footnote{$J_{\D_p}=p|\Omega\>\<\Omega|+(1-p)I/d$ has eigenvalue $dp+(1-p)/d$ on $|\Omega\>$ and $(1-p)/d$ elsewhere, and it is invariant under $\bar U\ts U$, so $\omega$ in~\eqref{eq:sdp} can be averaged to a multiple $cI$; the optimal $c=dp+(1-p)/d$ gives $\g^*=1+d^2p/(1-p)$.} Since $\D_{p'}$ is $\vareps$-EA-QDP, we have $p'\le p^*$ (trivially if $p'<0$), and therefore
\begin{equation}
  F(\E)\le F(\D_{p'})\le F(\D_{p^*})=\frac{\g+d-1}{\g+d^2-1},
\end{equation}
and likewise for $\hat T$, which proves~\eqref{eq:optimal-utility}. Finally, $\D_{p^*}$ is $\vareps$-EA-QDP, as the optimal $\omega=cI$ in the footnote is feasible with $\g^*=\g$, and it attains both bounds.
\hfill$\square$

\section{Privacy Parameters of Common Noise Mechanisms}\label{app:channels}

Table~\ref{tab:channel-values} lists the optimal QLDP parameter (=optimal QDP parameter with $\eta=1$), the optimal EA-QDP parameter, and the entanglement amplification of common single-qubit quantum channels. The QLDP values follow from~\cite{guan2025Optimal}.

\begin{table*}[!t]
  \centering
  \caption{Privacy Parameters of Common Single-Qubit Noise Mechanisms}
  \label{tab:channel-values}
  \small
  \renewcommand{\arraystretch}{1.35}
  \begin{tabular}{@{}lllccc@{}}
    \toprule
    Mechanism & Definition $\E(\rho)$ & Parameters & $\estar_\QLDP(\E)$ & $\estar_\EA(\E)$ & $A_{\rm ent}(\E)$ \\
    \midrule
    \multicolumn{6}{@{}l}{\emph{Pauli-type mechanisms} (Appendix~\ref{app:channels-pauli})}\\
    $X$/$Y$/$Z$ flip $\E_\sigma$ & $p\rho+(1-p)\,\sigma\rho\sigma$, \ $\sigma\in\{X,Y,Z\}$ & $0\le p\le1$ & $+\infty$ & $+\infty$ & --- \\
    \addlinespace
    $XY$/$YZ$/$XZ$ flip $\E_{\sigma\sigma'}$ & $p\rho+\frac{1-p}{2}(\sigma\rho\sigma+\sigma'\rho\sigma')$, & $0<p\le\frac13$ & $\ln\frac{1-p}{p}$ & $+\infty$ & $+\infty$ \\
     & \quad $\sigma\ne\sigma'\in\{X,Y,Z\}$ & $\frac13\le p<1$ & $\ln\frac{1+p}{1-p}$ & $+\infty$ & $+\infty$ \\
     & & $p\in\{0,1\}$ & $+\infty$ & $+\infty$ & --- \\
    \addlinespace
    Depolarizing $\D_p$ & $p\rho+(1-p)\frac I2$ & $0\le p<1$ & $\ln\frac{1+p}{1-p}$ & $\ln\frac{1+3p}{1-p}$ & $\frac{1+3p}{1+p}$ \\
     & & $p=1$ & $+\infty$ & $+\infty$ & --- \\
    \addlinespace
    General Pauli $\E_{\v w}$ & $\sum_{\alpha}w_\alpha\sigma_\alpha\rho\sigma_\alpha$, \ $\sigma_\alpha\in\{I,X,Y,Z\}$ & $w_{\min}>0$ & $\ln\frac{1+s}{1-s}$ & $\ln\frac{w_{\max}}{w_{\min}}$ & $\frac{w_{\max}}{w_{\min}}\cdot\frac{1-s}{1+s}$ \\
     & \quad ($s$: Eq.~\eqref{eq:app-pauli-s}) & $w_{\min}=0$, $s<1$ & $\ln\frac{1+s}{1-s}$ & $+\infty$ & $+\infty$ \\
     & & $s=1$ & $+\infty$ & $+\infty$ & --- \\
    \midrule
    \multicolumn{6}{@{}l}{\emph{Damping-type mechanisms} (Appendix~\ref{app:channels-damping})}\\
    Phase damping $\E_{\rm PD}$ & Eq.~\eqref{eq:app-pd} & $0\le\g\le1$ & $+\infty$ & $+\infty$ & --- \\
    \addlinespace
    Amplitude damping $\E_{\rm AD}$ & Eq.~\eqref{eq:app-ad} & $0\le\g<1$ & $+\infty$ & $+\infty$ & --- \\
     & & $\g=1$ & $0$ & $0$ & $1$ \\
    \addlinespace
    Generalized amplitude & Eq.~\eqref{eq:app-gad} & $0<\g<1$, $0<q<1$ & $\ln\frac{1+v}{1-v}$ & $\ln\frac{1+u}{1-u}$ & $\frac{(1+u)(1-v)}{(1-u)(1+v)}$ \\
    \quad damping $\E_{\rm GAD}$ & \quad ($u$, $v$: Eq.~\eqref{eq:app-gad-uv}) & $\g=1$ & $0$ & $0$ & $1$ \\
     & & $\g=0$; or $\g<1$, $q\in\{0,1\}$ & $+\infty$ & $+\infty$ & --- \\
    \bottomrule
    \multicolumn{6}{@{}l}{\footnotesize $w_{\max}:=\max_\alpha w_\alpha$ and $w_{\min}:=\min_\alpha w_\alpha$. ---: undefined, as $A_{\rm ent}$ is defined only when $\estar_\QLDP<+\infty$.}
  \end{tabular}
\end{table*}

\subsection{Pauli-Type Mechanisms}\label{app:channels-pauli}

To unify the $X/Y/Z$ flip channels, the depolarizing channel and other Pauli-type mechanisms, we define the general Pauli mechanism $\E_{\v w}$ with weights $w_\alpha\ge0$, $\sum_\alpha w_\alpha=1$, as
\begin{equation}
  \E_{\v w}(\rho)=\sum_{\alpha\in\{I,X,Y,Z\}}w_\alpha\,\sigma_\alpha\rho\sigma_\alpha.
  \label{eq:app-pauli}
\end{equation}
The other Pauli-type mechanisms are special cases of $\E_{\v w}$: $\E_\sigma$ has $w_I=p$ and $w_\sigma=1-p$; $\E_{\sigma\sigma'}$ has $w_I=p$ and $w_\sigma=w_{\sigma'}=(1-p)/2$, with $\E_{XZ}$ of Sec.~\ref{sec:motivating-example} at $p=1/3$; and $\D_p$ has $w_I=(1+3p)/4$ and $w_X=w_Y=w_Z=(1-p)/4$. Writing $\rho=\frac12(I+r_XX+r_YY+r_ZZ)$, the mechanism~\eqref{eq:app-pauli} maps $r_\alpha\mapsto s_\alpha r_\alpha$, with
\begin{equation}
  \begin{aligned}
    s_X&=w_I+w_X-w_Y-w_Z,\\
    s_Y&=w_I-w_X+w_Y-w_Z,\\
    s_Z&=w_I-w_X-w_Y+w_Z,
  \end{aligned}
  \quad
  s:=\max_{\alpha\in\{X,Y,Z\}}|s_\alpha|.
  \label{eq:app-pauli-s}
\end{equation}
In particular, $s=1$ for $\E_\sigma$, $s=\max\{p,1-2p\}$ for $\E_{\sigma\sigma'}$, and $s=p$ for $\D_p$. $\E_{\v w}$ is a unital channel, so all Pauli-type mechanisms are also unital. 

\subsection{Damping-Type Mechanisms}\label{app:channels-damping}

Besides Pauli-type mechanisms, another type of channels are the \emph{damping-type mechanisms}, which pull the qubit towards its ``ground state'' $|0\>$ and are not unital in general.
Let $\g\in[0,1]$ be the damping parameter\footnote{Note $\gamma$ here is the channel parameter, not to confuse with the objective variable in Theorem~\ref{thm:cp-sandwich}.}, and the damping-type channels are defined as follows:

\emph{Phase damping mechanism:}
\begin{equation}
  \begin{gathered}
    \mathcal{E}_{\mathrm{PD}}(\rho) = K_0 \rho K_0^\dagger + K_1 \rho K_1^\dagger,\\
    K_0=\begin{bmatrix}1&0\\0&\sqrt{1-\g}\end{bmatrix},\quad
    K_1=\begin{bmatrix}0&0\\0&\sqrt{\g}\end{bmatrix}.
  \end{gathered}
  \label{eq:app-pd}
\end{equation}
It equals $\E_Z$ with $p=(1+\sqrt{1-\g})/2$.

\emph{Amplitude damping mechanism:}
\begin{equation}
  \begin{gathered}
    \mathcal{E}_{\mathrm{AD}}(\rho) = K_0 \rho K_0^\dagger + K_1 \rho K_1^\dagger,\\
    K_0=\begin{bmatrix}1&0\\0&\sqrt{1-\g}\end{bmatrix},\quad
    K_1=\begin{bmatrix}0&\sqrt{\g}\\0&0\end{bmatrix}.
  \end{gathered}
  \label{eq:app-ad}
\end{equation}
For $\g=1$, it is the replacement map $\rho\mapsto|0\>\<0|$.

\emph{Generalized amplitude damping mechanism,} with ground-state probability $q\in[0,1]$ of the thermal state $q|0\>\<0|+(1-q)|1\>\<1|$:
\begin{equation}
  \begin{gathered}
    \begin{aligned}
      \mathcal{E}_{\mathrm{GAD}}(\rho)
      &=K_0 \rho K_0^\dagger + K_1 \rho K_1^\dagger
      +K_2 \rho K_2^\dagger + K_3 \rho K_3^\dagger,
    \end{aligned}\\
    \begin{aligned}
      K_0&=\sqrt q\begin{bmatrix}1&0\\0&\sqrt{1-\g}\end{bmatrix}, &
      K_1&=\sqrt q\begin{bmatrix}0&\sqrt\g\\0&0\end{bmatrix},\\
      K_2&=\sqrt{1-q}\begin{bmatrix}0&0\\\sqrt\g&0\end{bmatrix}, &
      K_3&=\sqrt{1-q}\begin{bmatrix}\sqrt{1-\g}&0\\0&1\end{bmatrix}.
    \end{aligned}
  \end{gathered}
  \label{eq:app-gad}
\end{equation}
It reduces to $\E_{\rm AD}$ for $q=1$. For $0<\g<1$ and $0<q<1$, the parameters in Table~\ref{tab:channel-values} use
\begin{equation}
  u:=\frac{\sqrt{1-\g}}{\sqrt{ab}},\qquad
  v:=\frac{\sqrt{1-\g}}{\sqrt{1-\g(2q-1)^2}},
  \label{eq:app-gad-uv}
\end{equation}
where $a:=1-\g+q\g$ and $b:=1-q\g$, so that $ab=1-\g+q(1-q)\g^2$. Here $a$ and $b$ are the $|00\>$ and $|11\>$ diagonal entries of $J_{\E_{\rm GAD}}$, whose only off-diagonal entries are $\sqrt{1-\g}$ between $|00\>$ and $|11\>$; $\omega=(1+u)\,\mathrm{diag}(a,b)$ is optimal in~\eqref{eq:sdp}.

\end{CJK*}
\end{document}